# Knowledge Emergence in Scientific Communication:
# From "Fullerenes" to "Nanotubes"

Diana Lucio-Arias & Loet Leydesdorff
Amsterdam School of Communications Research (ASCoR)
Kloveniersburgwal 48, 1012 CX Amsterdam, The Netherlands

**Abstract**

This article explores the emergence of knowledge from scientific discoveries and their effects on the structure of scientific communication. Network analysis is applied to understand this emergence institutionally as changes in the journals; semantically, as changes in the codification of meaning in terms of words; and cognitively as the new knowledge becomes the emergent foundation of further developments. The discovery of fullerenes in 1985 is analyzed as the scientific discovery that triggered a process which led to research in nanotubes.

## 1. Introduction

New frontiers of scientific research provide us with one of the best environments to explore the process of knowledge production from the perspective of scientific communication and its diffusion to other societal domains. This paper addresses questions such as how existing knowledge converges into new discoveries, how these discoveries are transmitted through scientific communication, and how they become knowledge (meaningful information) that gives way to new scientific developments and discoveries. Time-dependent analysis is required to understand this transition from information to meaningful information to knowledge.

Development of nanoscale research and the attention it has captured due to its potential applications and implications suggest a case for the study of the dynamics inherent to the emergence and consolidation of research fronts. In nanoscience, fields of research like chemistry, physics, computer science, and biology, merge in the study of nanometer[1] scale structures (equivalent in size to materials in the range from hundreds to tens of nanometers in size) and their properties.

Although research at the nanoscale can be dated back to the 19th century with the appearance of colloidal and surface science, it was not until the invention of high-resolution electron microscopes, particularly the Scanning Tunneling Microscope, that research at this level could be subject of scientometric study because a field could increasingly be distinguished. In 1997 the rate of growth of scientific journal papers with the prefix “nano” in their titles was exponentially doubling over time (Braun *et al.,* 1997 and Meyer & Persson, 1998). This exponential growth has persisted aided by increased attention of science and technology policies for “nano” (Schummer, 2004). Scientometric studies of the field have grown accordingly.

This study focuses on research in “fullerenes” and the fullerene-like structures “nanotubes.” This field of research originated after the discovery of fullerenes in 1985 from the study of properties of carbon molecules, especially the $C_{60}$ molecule. Research about “fullerenes” can be recognized as a significant research front of nanoscience, having developed its own terminology beyond the usage of the prefix “nano” (Schummer, 2004). A series of events can be placed along the “evolution” of this research front: a Nobel Prize was awarded for the discovery of fullerene in 1996; the journal *Fullerene, Science and Technology* was launched in 1994; the change of the name of this journal in 2002 to *Fullerene, Nanotubes and Carbon Nanostructures*; and the discovery in 1991 of cylindrical one-dimensional fullerenes, later named carbon nanotubes, with potential uses in very diverse areas such as disease treatment, computer memory, methods of drug delivery, cosmetics, alternative uses of energy, etc.

## 2. On scientific communication

New knowledge can be considered as codification of communication that disrupts the state of the art of existing knowledge. In this paper, emphasis on scientific communication will be made to track emergence of knowledge generated from new scientific discoveries. By concentrating on scientific texts, knowledge production can be made the subject of measurement considering the organizational, textual and cognitive dimensions of scientific developments (Leydesdorff, 1995). Knowledge production is structurally coupled to the circulation of texts (Fujigaki, 1998). These texts are: (a) institutionally arranged in scientific journals produced by individuals affiliated to institutions and organizations, (b) semantically structured to carry meaning, and (c) cognitively associated to the existing body of knowledge through links formed by citations (*Ibid.*).

[1] Nanometer = $10^{-9}$meters = a billionth of a meter

Producing and publishing scientific texts is one of the main activities of scientists. Peer reviewing of produced texts prior to publication validates the scientists' work. Issues related to money allocation, prestige, and a scientist's field of work can also be defined in relation to the output of publications. Through publishing, scientists select validated knowledge and differentiate their own results and conclusions to submit them for further validation. This is a continuous process triggered by its own self-referentiality (as a process). The recursive process of selections and codifications carries the development of knowledge (Fujigaki, 1998).

Journals aggregate validated scientific knowledge represented in texts. Journals can be constituted where the codes of communication that the scientists use for communicating their results can be stabilized (Leydesdorff, to be published). These codes are further differentiated and respond to characteristics such as field or topic specificity, country or language. Aggregation of these differentiated codes of communication forms the journal space. One can expect that the differentiation can only be reproduced if it is also functional for the further development of knowledge.

Different journals are related in the journal space through a network of cited and citing relations. New discoveries generate new meaning that can irritate and change this network. These changes may imply the emergence of a new scientific journal that specializes in the communication generated by the discovery and its consequent knowledge or—if less fundamental—it can be absorbed in existing structures.

Variation generated by new discoveries can be measured through a change in the state of relations in the journal space. However, because knowledge is a semantic (and social) construct, variation from new discoveries can also be measured in relation to changes in the semantic representation of the social construction of knowledge.

Knowledge emerges semantically in a network of relations among words selected to carry meaning (Hellsten & Leydesdorff, 2005). This process of selection generates a network-structure of co-words where the position of the words is defined by the presence and absence of relations among them (Hesse, 1980; Burt, 1982). Thus, meaning and knowledge are intrinsic characteristics of the network: while meaning is easily measurable in the specific configuration of the network, knowledge requires a further analysis considering the time axis. Knowledge is a further codification of meaning, where meaning is information which is already codified.

These codifications depend on the purpose of the communication; meaning can be asymmetrically codified across domains. These asymmetries are caused by the different interpretations of meaning carried by words and co-occurrences of words across domains (Leydesdorff & Hellsten, 2005). In this study, the emergence of knowledge originating from a discovery will be considered as: (a) variation in the codes of scientific communication, (b) variation in the semantic structure of knowledge representations at different moments in time (c) further codification of the potentially serendipitous[2] discovery from previously constructed knowledge. An analysis of the codification in patents will be briefly presented as an expansion into knowledge-based applications assuming that these are a consequence of the knowledge constructed through scientific communications.

Our particular interest resides in the question of how the codification of meaning changes with time resulting in changes in the communication structures and how different meaning codifications outlay the emergence of new knowledge that builds on –but differentiates from– the existing body of knowledge.

## 3. Methodology

To understand structural changes derived from the emergence of new knowledge we use methods from network analysis at various levels to: (a) track changes in the network of relations of journals, (b) measure changes of meaning through changes in the relations of words at different times and in different domains (patents), and (c) understand the construction of knowledge as a process of selecting and differentiating from previous knowledge.

[2] Carbon nanotubes were discovered by Sumio Iijima, who refers to this finding as a serendipitous result of research in fullerenes.

### *3.1 Journals*

The journal space (that is, the aggregation of publications in scientific journals) defines a cognitive structure that differs from other journal spaces and contextualizes papers belonging to it (Cozzens, 1985). Network analysis allows us to pinpoint journal spaces by assuming that the citation relations are cognitive relations as well (Amsterdamska & Leydesdorff, 1989). A journal space is composed of journals that have citation relations between them.

Definition of the journal space allows limiting the environment of a specific journal identifying the journals from which it "feeds". The delineation of the environment of a journal is a relational problem that will be addressed considering the relevant environment in the overall database of citations taken from the *Journal Citation Reports* of the *Science Citation Index*. The mapping can be complemented with a factor analysis approach in order to outlay the position of the journal in a network of relations and its underlying dimensions (Leydesdorff & Cozzens, 1993).

For the special case of fullerenes research, the appearance of the journal *Fullerene, Science and Technology* can be considered as an important breakthrough, since the journal specialized in scientific communication related to fullerene issues. The journal's environment was delineated by considering all citations to and from this journal. The citation environment of this journal will be studied for the years 1996 (the first year in which it appeared in the *Journal Citation Reports*) and 2004 in order to consider changes in the construction of boundaries.

Normalization using the cosine as proposed by Salton & McGill (1983) will be applied to the journals environment defined by the matrix of citation relations. Visualizations are optimized using the algorithm of Kamada & Kawai (1989)

### *3.2 Codification of meaning in time and space*

The study of scientific developments through the network of relations among words in texts has already been proposed by Hesse (1980) and Callon *et al*. (1986). In actor-network theory, the strength of the relations of co-words depends on the network of associations that form its context (Theil & Latour, 1995). The semantic maps in this study, however, are not derived from the relations between two specific words but among any number of words in a set of titles. For this, the asymmetric matrix of words versus the retrieved titles of documents was used. The relations were normalized using the cosine among word vectors as proposed by Salton & McGill (1983). This provides use with spatial representations of how words are positioned in relation to other words.

Two different sources of information were considered to track changes of meaning in domains: *the Science Citation Index (SCI)* as available in the so-called Web of Science (at http://scientific.thomson.com/products/wos/); and the patents search facility made available by the *United* States Patent Office (USPTO) (http://www.uspto.gov/patft/). In both databases, records that matched the criteria "fullerene" (or "fullerenes") and "nanotube" (or "nanotubes") as title words were retrieved until and including 2005.

The study of the influence of time in the codification of meaning is explored only for scientific communication (i.e., the *SCI*) because the sets retrieved from the USPTO are too small for this purpose. Given that time is not relevant if it is not related to a series of events that may introduce variation to the codification of meaning, the number of published documents was considered as the clock-time. Initially, the analysis was made with pragmatic cut-offs for every 1,000 documents. However, for the purpose of this paper, three different moments for each set were selected with equal distances in terms of numbers of events, that is, each 2,565[3] documents in the case of fullerene(s), and 3,224 documents in the case of nanotubes(s).

The asymmetric matrix of words versus titles was obtained using TI.exe (at: http://www.leydesdorff.net/software.htm ) that generates a word-occurrence matrix, and a normalized co-occurrence matrix from a set of titles and a word list. The matrices were then read using the software package for network analysis PAJEK (at: http://vlado.fmf.uni-lj.si/pub/networks/pajek/). All the visualizations will be based on the algorithm of Kamada & Kawai (1989), present in PAJEK. For the sets retrieved from the *SCI* only words that occur more than 20 times, will be considered and the cosine is set at $\geq$ 0.2 in order to optimize the visualizations. For the case of the patents titles, all words occurring more than two times are used and the visualizations are optimized using only the cosine-normalized relations stronger than 0.3.

*3.3 Citations*

[3] The 7,696 documents retrieved for fullerene were divided into two sets of 2,655 and the last set was of 2,656 documents

In this paper we use citation analysis to delineate the environment of a journal, as explained previously. Citations inside the set of documents retrieved are also interesting as they can be used to track the history of the development of knowledge embedded in them.

Citation relations do not always represent the scientist's use of results previously achieved. Nevertheless, through citations scientists link their work to the existing literature generating a network structure that can be used to re-write the history of scientific developments (Garfield, 1973).

Citation behaviors distinguish two different types of scientific literature: classic literature that have high citation rates independently of their age; and transient literature that becomes obsolete short after their publication and looses citing importance (Chen, 2006). In the first case, the literature belongs to the validated knowledge-base while the second type of literature is typical of the rapidly changing dynamics of the research fronts (*Ibid.*).

For analyzing the citing relations inside our set of documents, the software package *HistCite*™ (http://www.histcite.com/) was used. *HistCite* allows to build citation histographs considering the retrieved documents from the *SCI*. The documents are represented by nodes, displayed historically and their size depends on the amount of citations it gets inside the set.

## 4. Results

For the case of fullerene, a search in the *SCI* retrieved 7,696 documents with "fullerene(s)" in their titles. The first document of the set was published in *Nature* by Harold W. Kroto, one of the three scientists awarded with the Nobel Prize for the discovery of fullerenes in 1996. This document was titled: "The Stability of the Fullerenes C-24, C-28, C-32, C-36, C-50, C-60 and C-70" (Kroto, 1987). Of the total set of documents, 84% were articles, 5% meeting abstracts, and 11% either reviews, letters, notes or editorial matter. For the second case, 9,672 documents were downloaded from the *SCI* with "nanotube(s)" in their title; 87% of them were scientific articles. The earliest document was entitled "Large Scale Synthesis of Carbon Nanotubes" (Ebbesen, *et al.*, 1992) and published in *Nature*.

The evolution of research in fullerenes is depicted in Figure 1 as the annual growth of documents having "fullerene(s)" or "nanotube(s)" as words in their titles. Notable events such as the emergence of the journal, the Nobel Prize awarded to Harold W. Kroto, Richard E. Smalley and Robert F. Curl Jr. for the discovery of fullerenes, and the discovery of the one-dimensional fullerenes, that were named carbon nanotubes, are also represented in this figure.

**Figure 1**. Number of documents in the *SCI* with "fullerene" and "nanotube" in their titles

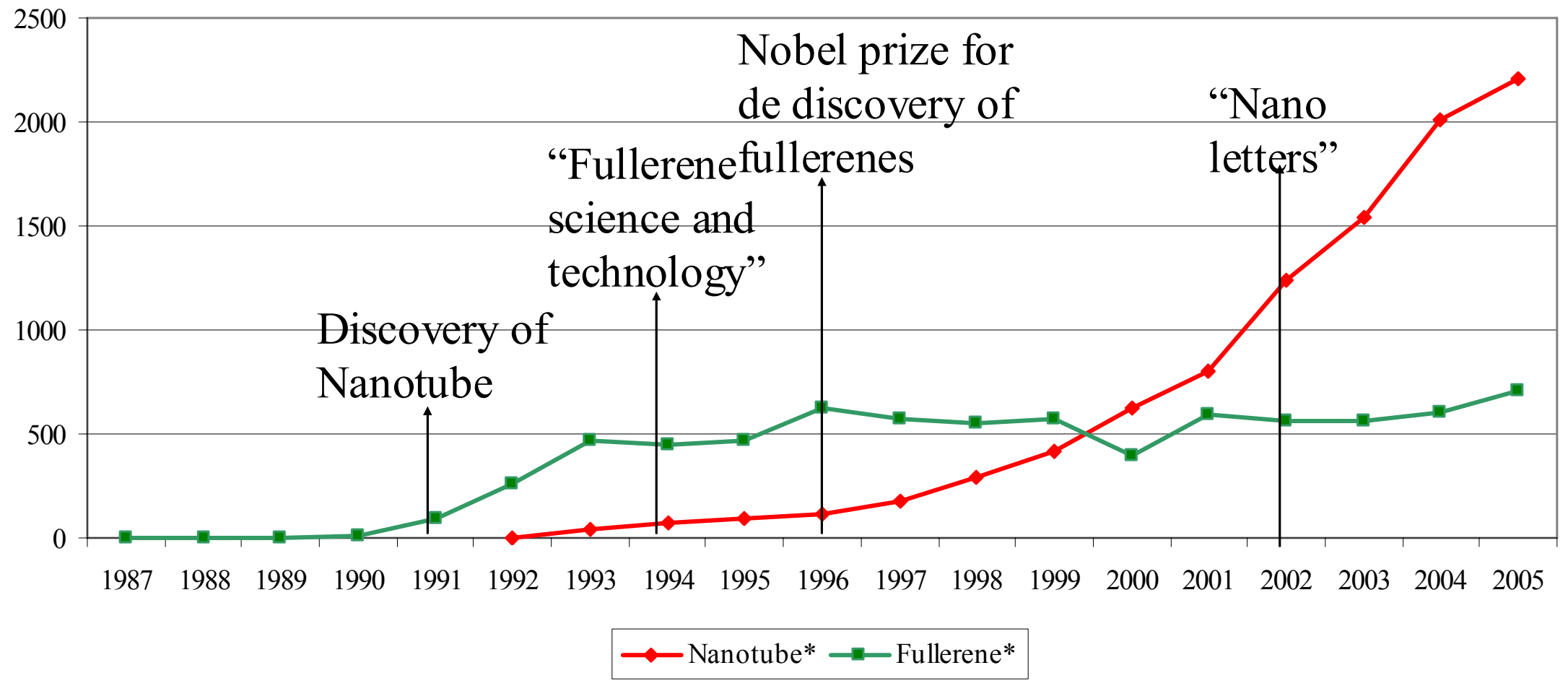

over time.

*4.1 Journal differentiation*

The system of scientific communication is highly differentiated; it achieves organization through the functional differentiation of journals that allow the communication of more complexity (Cozzens & Leydesdorff, 1993). Journals are positioned in the journal space in accordance to the relations they have with other journals; this positioning allows the delineation of scientific specialties as clusters of scientific journals (*Ibid.*). New discoveries produce variation that disrupts the organization of scientific communication. For example, the emergence of a new journal can respond to specialization of the communication because of the emergence of new research fronts.

Although for both sets of documents—those retrieved with "fullerene(s)" and those with "nanotube(s)" in their title—the distribution in traditional chemistry and physics journals is similar to those found in other studies about nanoscience research (Meyer & Persson, 1998), a significant number of the documents containing "fullerene(s)" in their title words is published in *Fullerene, Science and Technology.* In Figure 2, the distribution of the journals publishing the 9,769 documents retrieved for "fullerene(s)" in their title is shown. Although *Chemical Physics Letter* published more documents, the share of fullerene-related articles in each journal is highest for *Fullerene, Science and Technology* (72% of the total number of articles).

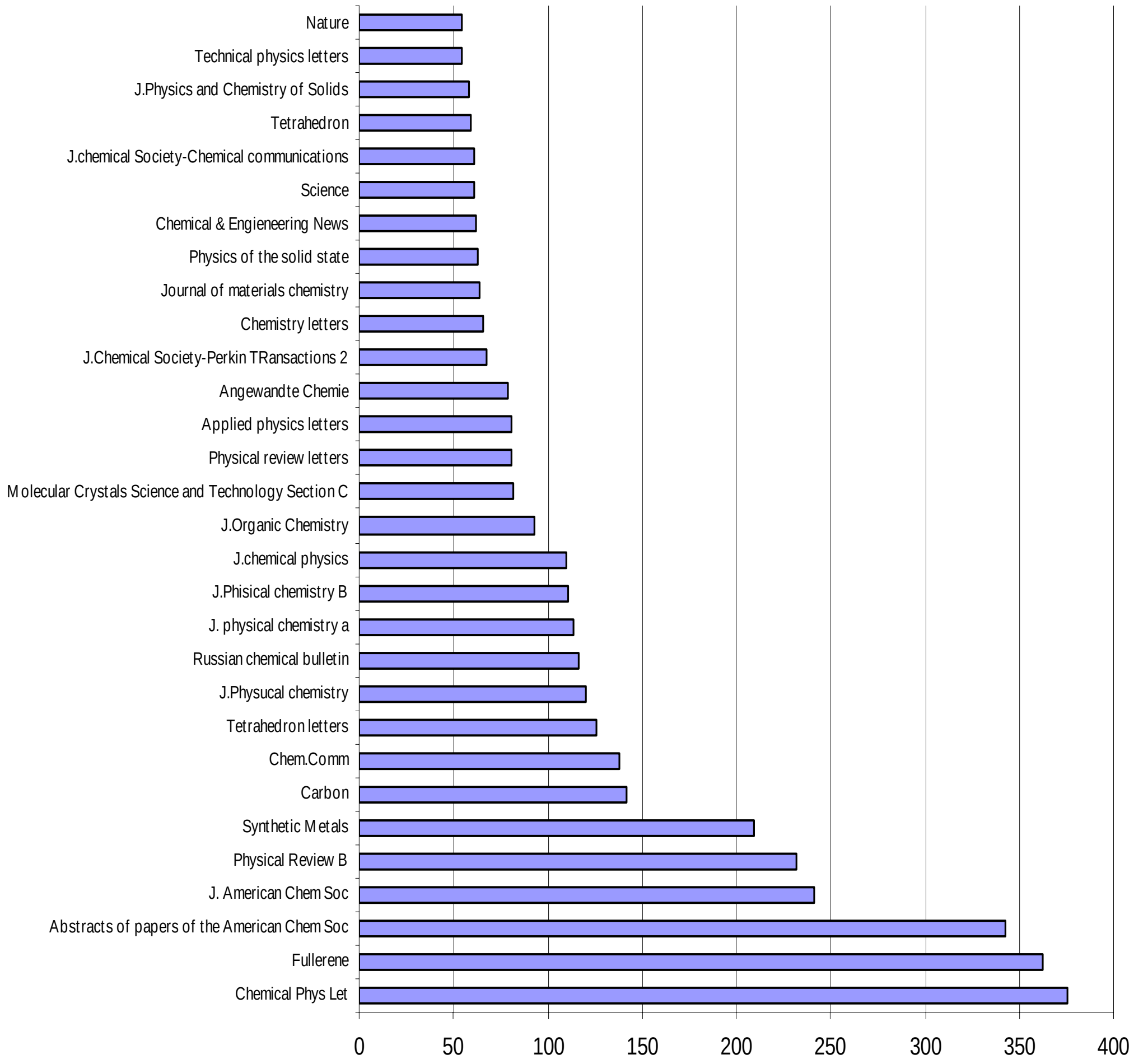

**Figure 2**. Distribution of fullerene-related documents in journals, 1987-2005

*Fullerene, Science and Technology* changed its name in 2002 into *Fullerene, Nanotubes and Carbon Nanostructures.* This journal has published 840 documents of which 92% are classified as scientific articles.

In Figure 3, the evolution of the number of documents published in the journal is shown. Although it is not a high number of documents and, as will be seen further, it is not highly-cited, the constitution of the journal can be considered as very important in the organization of communication in the field. Further analysis of the journal's contents and authors could give more insight on the importance of the emergence of the journal as self-organization of the communication in the nano-field.

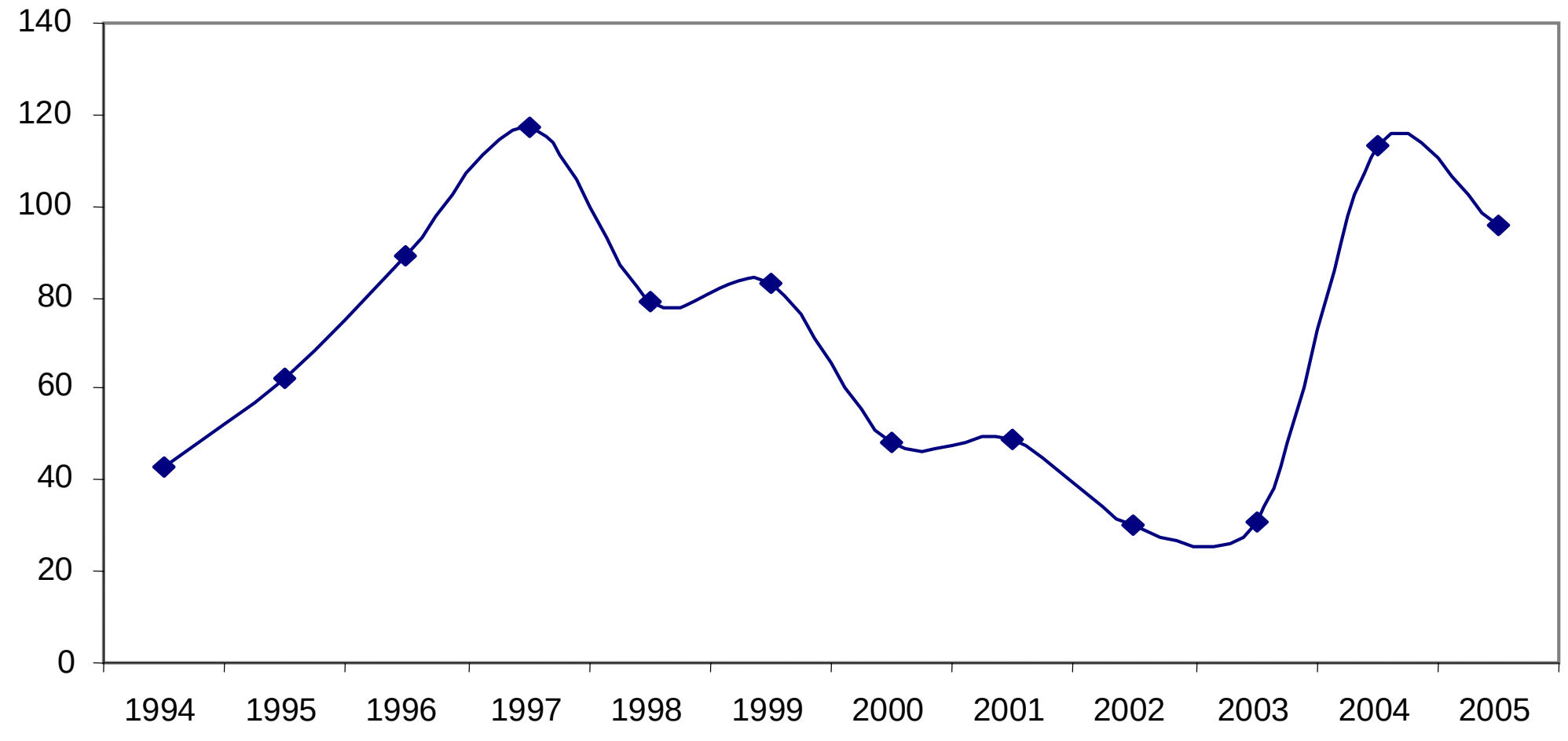

**Figure 3**. Number of documents in the journal *Fullerene, Science and Technology* (since 2002: *Fullerene, Nanotubes and Carbon Nanostructures*).

Two different network approaches can be applied to the position of a journal in a set of citations to other journals. The first analysis uses the cited journals to understand the position of a specific journal among a series of established cognitive structures. For this specific case, an analysis of the journals cited in *Fullerene, Science and Technology* is first presented for 1996, the year in which the journal appears for the first time in the *Journal Citation Reports*. This result is analyzed similarly to the environment for the year 2004 in order to illustrate changes in its boundaries ten years after its constitution. The second analysis uses the other journals citing *Fullerene Nanotubes and Carbon*

*Nanostructures* for the year 2004 when the definition of the journal's boundaries is already set.

For the year 1996, the matrix representing the citation environment of *Fullerene, Science and Technology* consisted of 18 cited journals. The factor analysis of this citation environment is shown in Table 1.

**Rotated Component Matrix(a)**

| | Component | | | |
|---|---|---|---|---|
| | 1 | 2 | 3 | 4 |
| PHYS REV B | .948 | | | |
| SOLID STATE COMMUN | .947 | | | |
| APPL PHYS A-MATER | .879 | | -.102 | .113 |
| JPN J APPL PHYS | .308 | -.241 | -.180 | .284 |
| CARCINOGENESIS | -.210 | -.188 | -.105 | .167 |
| CHEM PHYS LETT | | .920 | | |
| J PHYS CHEM-US | | .906 | | .105 |
| INT J QUANTUM CHEM | | .814 | | |
| CROAT CHEM ACTA | -.172 | .284 | | |
| B SOC CHIM BELG | | | .899 | |
| J ORG CHEM | | | .876 | |
| **FULLERENE SCI TECHN** | | **-.116** | **-.119** | **-.806** |
| TRANSIT METAL CHEM | | | | -.664 |
| J CRYST GROWTH | | -.253 | -.151 | .349 |
| J ELECTROANAL CHEM | -.151 | .227 | | .266 |
| SYNTHETIC MET | .219 | .113 | | .251 |
| J PHYS B-AT MOL OPT | | | -.174 | -.219 |
| CHEM VAPOR DEPOS | -.191 | | -.129 | .206 |

Extraction Method: Principal Component Analysis.
Rotation Method: Varimax with Kaiser Normalization.
a Rotation converged in 6 iterations.

**Table 1**. Factor structure among 18 journals citing *Fullerene, Science and Technology* in 1996

Table 1 shows the organization of the 18 journals which cite *Fullerene, Science and Technology* in 1996, in four dimensions explaining 49.2% of the variance. The first three components are physics, physical chemistry, and organic chemistry. *Fullerene, Science and Technology* has low or negative loadings on these factors, but leads a fourth factor as a separate grouping. *Transition Metal Chemistry* is also attributed to this group.

In Figure 4, the citation environment just discussed with a factor analytical approach is visualized.

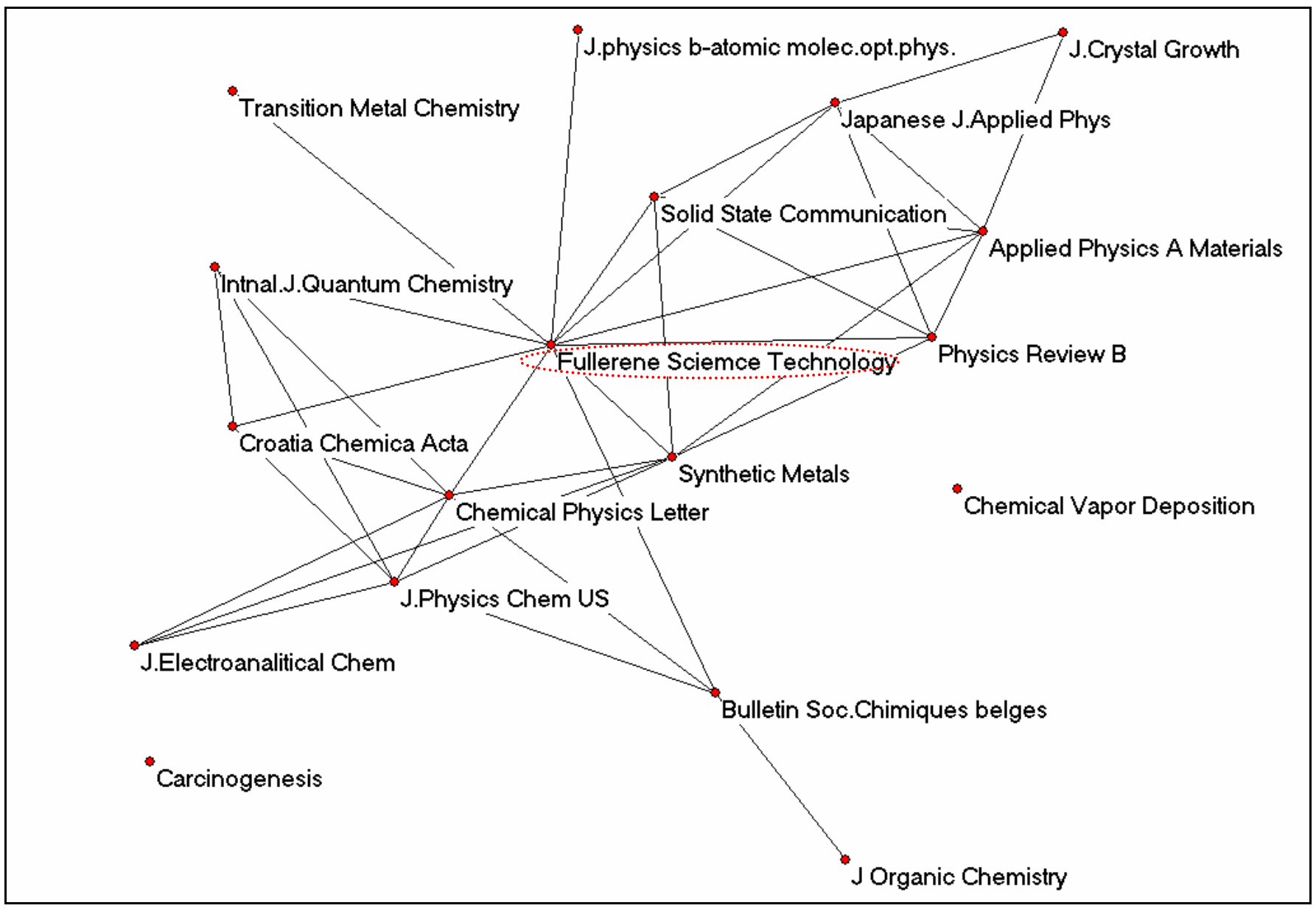

**Figure 4**. Eighteen journals citing *Fullerene, Science and Technology* in 1996 (cosine ≥ 0.2)

The same exercise was repeated for 2004 in order to illustrate possible changes in the disciplinary foundations of fullerene-related research. In 2004, *Fullerene, Nanotubes and Carbon Nanostructures* is cited by only 14 other journals.

**Rotated Component Matrix(a)**

| | Component | | | |
|---|---|---|---|---|
| | 1 | 2 | 3 | 4 |
| CHEM PHYS LETT | .900 | | | |
| CHEM PHYS | .876 | | -.110 | |
| PHYS CHEM CHEM PHYS | .826 | | | |
| J PHYS CHEM B | .805 | | .163 | |
| TETRAHEDRON | -.144 | .762 | .118 | -.116 |
| J FLUORINE CHEM | -.180 | .640 | -.207 | .116 |
| ANGEW CHEM INT EDIT | | .551 | .367 | -.324 |
| RUSS J INORG CHEM+ | -.265 | -.307 | .162 | |
| RADIAT PHYS CHEM | | -.135 | -.604 | -.140 |
| POLYHEDRON | -.221 | -.230 | .603 | -.107 |
| J MATER CHEM | .390 | | .499 | |
| ORIGINS LIFE EVOL B | -.180 | -.244 | -.179 | -.666 |
| **FULLER NANOTUB CAR N** | | **-.202** | | **.566** |
| CARBON | | | | .523 |

Extraction Method: Principal Component Analysis.

Rotation Method: Varimax with Kaiser Normalization.

a Rotation converged in 6 iterations.

**Table 2**. Factor structure among 14 journals citing *Fullerene, Nanotubes and Carbon Nanostructures* in 2004.

In 2004, four factors explain 52.4% of the variance in the matrix. The physics environment has disappeared. *Fullerene, Nanotubes and Carbon Nanostructures* forms a fourth factor with the journal *Carbon.* It is worth considering that from 2001 to 2003 the total number of documents published in the journal revealed a strong descent (see Figure 3) that influence the overall cited environment of the journal.

Figure 5 visualizes the citation environment that was factor analyzed in Table 2 on the basis of the cosine as the similarity measure. This visualization shows that a strong component of the graph is formed by six journals in the field of physical chemistry and materials chemistry. *Fullerene, Nanotubes and Carbon Nanostructures* is one of these six journals.

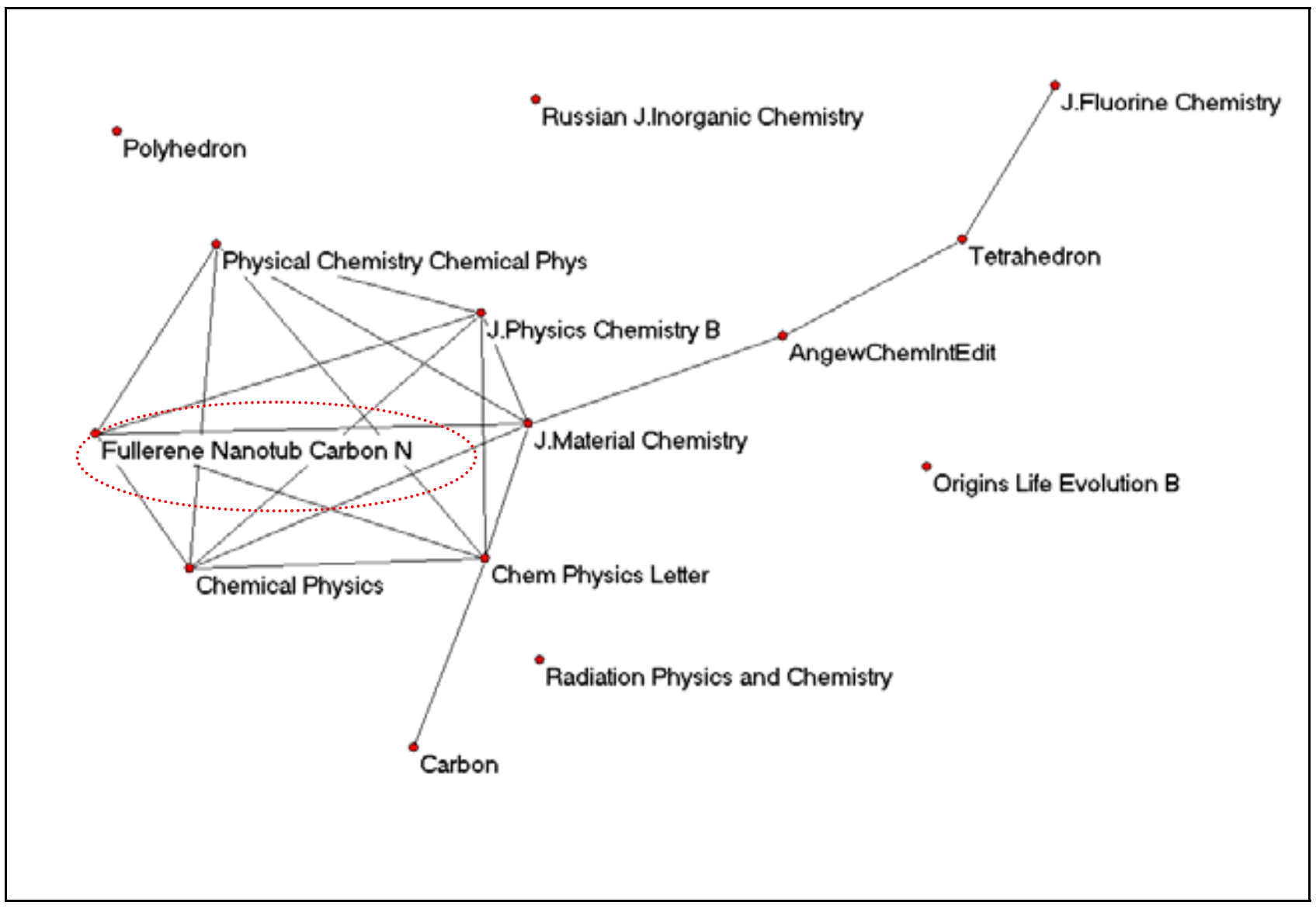

**Figure 5**. Fourteen journals citing *Fullerene, Nanotubes and Carbon Nanostructures*, 2004 (cosine ≥ 0.2)

It is also important to consider how authors in the journal *Fullerene, Nanotubes and Carbon Nanostructures* see the environments when citing other journals. The previous exercise for the journals citing *Fullerenes Nanotubes and Carbon Nanostructures* was also done for the journals cited by it in 2004. These are 95 journals. To reduce the amount of variation in the network due to the volume of journals that are infrequently cited, just the journals that constitute more than one percent of the total references is considered (He & Pao, 1986; Leydesdorff & Cozzens, 1993). Fifteen journal are cited to the extent of more than one percent of the total number of 1,122 references provided by 104 articles published in this journal in 2004

**Rotated Component Matrix(a)**

| | Component | | | |
|---|---|---|---|---|
| | 1 | 2 | 3 | 4 |
| CHEM COMMUN | .912 | -.204 | | |
| J AM CHEM SOC | .888 | | .273 | |
| J ORG CHEM | .872 | -.176 | -.115 | -.172 |
| ANGEW CHEM INT EDIT | .816 | -.158 | | |
| RUSS CHEM B+ | .787 | -.252 | | -.210 |
| TETRAHEDRON LETT | .756 | -.192 | -.264 | -.208 |
| PHYS REV B | -.119 | .892 | .246 | |
| PHYS REV LETT | -.114 | .788 | .193 | .184 |
| PHYS SOLID STATE+ | -.241 | .776 | | -.170 |
| APPL PHYS LETT | -.221 | .528 | | |
| **FULLER NANOTUB CAR N** | | **.197** | **.931** | |
| CHEM PHYS LETT | .136 | .214 | .846 | |
| CARBON | -.431 | -.251 | .486 | |
| SCIENCE | | | | .959 |
| NATURE | -.105 | | | .941 |

Extraction Method: Principal Component Analysis. Rotation Method: Varimax with Kaiser Normalization.
a Rotation converged in 6 iterations.

**Table 3**. Factor structure among 15 journals cited by *Fullerene, Nanotubes and Carbon Nanostructures, 2004*

Table 3 shows that for 2004, the 15 journals cited by *Fullerenes Nanotubes and Carbon Nanostructures* organized in four dimensions. These four dimensions explain 75.7% of the variance. The "self-"references within the journal correlate highly with the references to papers in *Chemical Physics Letters* and *Carbon.* These three journals load on a third factor which explains 13.2% of the variance. Unlike the relevant citation environment, authors publishing in these journals relate to a physics factor (17.1%) which is the second one when compared with a chemistry factor (35.8%).

*Nature* and *Science,* with a high load in the fourth component, can be considered as multidisciplinary; their appearance in the citation environment of *Fullerene, Nanotubes and Carbon Nanostructures* might be explained by the diffusion of new discoveries in multidisciplinary journals aimed at a broader audience. We shall see below that the articles which are most cited in this field appear in *Science* and *Nature.*

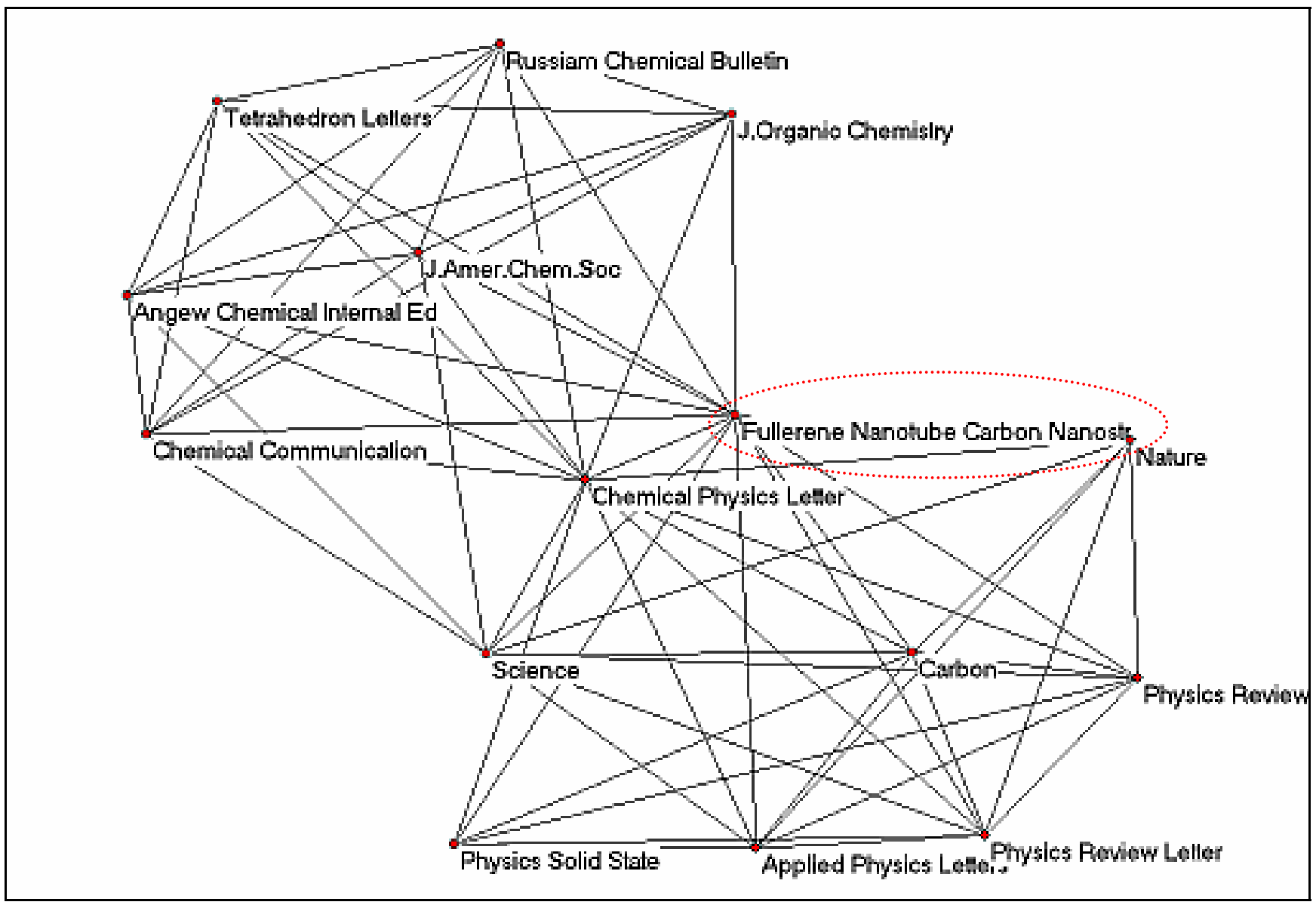

**Figure 6**. Fifteen journals cited within articles in *Fullerene, Nanotubes and Carbon Nanostructures* in 2004 (cosine ≥ 0.2).

*4.2 Codification of meaning in time and space*

Until now, the emergence of knowledge has been conceptualized in terms of variation in the structure of the journal space. New discoveries disrupt the network structure communicating scientific developments. In the following section, the evolution of the term "fullerene" will be analyzed considering semantic changes that occur in time and in a different domain that the one represented by scientific communication.

Nanotubes are fullerene-like structures. They were discovered in 1991. The fact that nanotubes have more potential applications than fullerenes may explain the exponential growth of documents with the word "nanotube" in their titles (Figure 1). Because nanotubes are a specific form of fullerenes, research with this focus is related to research in fullerenes. Nanotube discoverer, Sumio Iijima, defined his discovery of nanotubes as serendipitous while researching fullerene carbon structures (Ijima, 2002). Research in nanotubes as some specific types of fullerenes made us hypothesize that through the codification of the concept of "nanotube" a further codification of "fullerene" can be explained. For this reason, fullerene and nanotube related documents are both considered.

For the case of codification in a different space than the one represented by scientific communication, the codification from association of words in the titles of the patents of the *United States Patents and Trademark Office* (www.uspto.gov) was analyzed.

### *4.2.1 Time*

This part of the analysis emphasizes on the effect of time in the codification of the concept of fullerene and the posterior feedback-effect of further research in a specialized application of fullerenes (nanotubes). Documents with fullerene(s) or nanotube(s) in their titles were retrieved from the *SCI.* Because time is considered in this study as a sequence of events that produce a difference,[4] each of the sets was divided in three sets with an equal number of consecutively published documents. The purpose of this division was to find out how the codification of the concept changes and if this change could be observed as a change in the process of producing and validating knowledge.

For every case, only words occurring more than 20 times were considered, and the search terms (fullerene(s) and nanotube(s)) were excluded from the analysis. The cosine was used for the normalization. The algorithm of Kamada and Kawai (1989) available in the software PAJEK was applied to the visualization of the networks with a threshold of cosine $\geq 0.2$.

*Fullerene documents*

For the set of fullerenes, the division was made into two subsets of 2,565 and one of 2,566 documents. The first set, Figure 7, corresponds to documents published from 1987 to 1997. A total of 4,645 words occur in the titles of these documents, and 1,905 occur more than once. For this case 90 words that appeared more than once were considered.

[4] And not only a variable which may co-vary. For a more detailed explanation of the dynamic analysis of network structures: Leydesdorff, 1991.

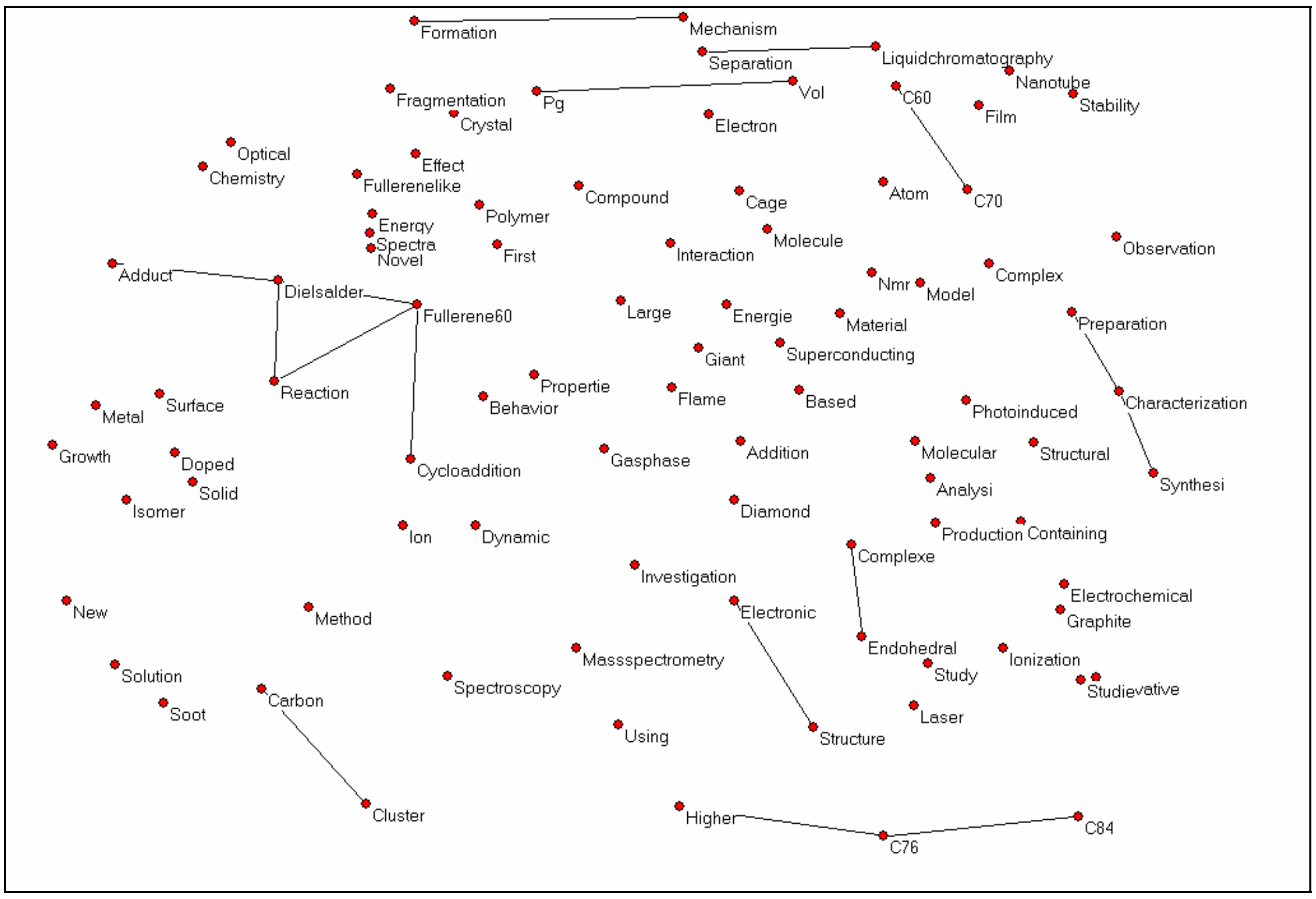

**Figure 7**. Cosine map of the first 2,565 titles in fullerene related documents. (cosine ≥ 0.2)

As can be seen from Figure 7, this first set depicts a loose set of nodes with very few clusters mostly composed of dyadic relations between words. The carbon nanotube, fullerene-like structures that were first spotted in 1991, have a high frequency of occurring inside the fullerene-related documents, but their relation vanishes when the threshold of the cosine is set at 0.2. In other words, the patterns of relations among the title words are not specific. The loose association of words illustrates a first stage of research dominated by documents related to the chemistry of fullerenes and process associated with the production and synthesis of fullerene-like structures.

For the second set, shown in Figure 8, the documents were published between 1997 and 2001 and the total number of words in these titles was 5,478 of which 2,238 happened at least twice. For the analysis, 127 words occurring more than 20 times were considered.

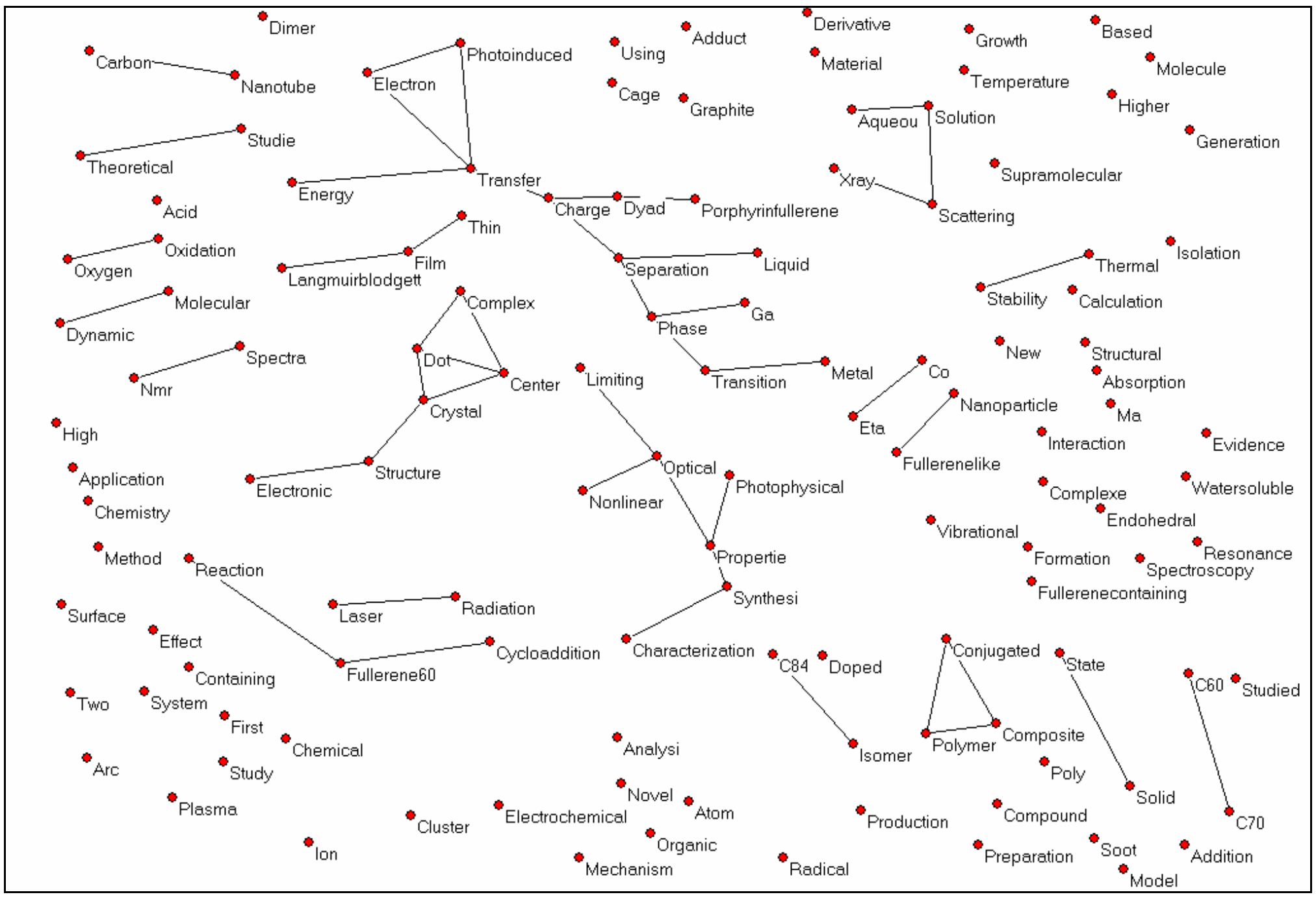

**Figure 8**. Cosine map of the second 2,565 titles in fullerene related documents. (cosine ≥ 0.2)

In Figure 8, an increase of the sophistication of the titles could be perceived from the denser network of relations among their words. With the threshold set at cosine ≥ 0.2, although the network is not yet very dense, the number and complexity of the clusters increases from the one observed in Figure 7. The carbon-nanotube link remains in this second set when the threshold increased illustrating a greater importance of nanotubes in fullerene-related research than in the earlier documents. It is in the period here considered that nanotube related documents start to grow exponentially.

The last set, shown in Figure 9, comprised documents published between 2001 and 2005; in these titles, 5,102 words appear and 2,165 appeared at least two times. The analysis was made with 144 words that appeared more than 20 times in the set.

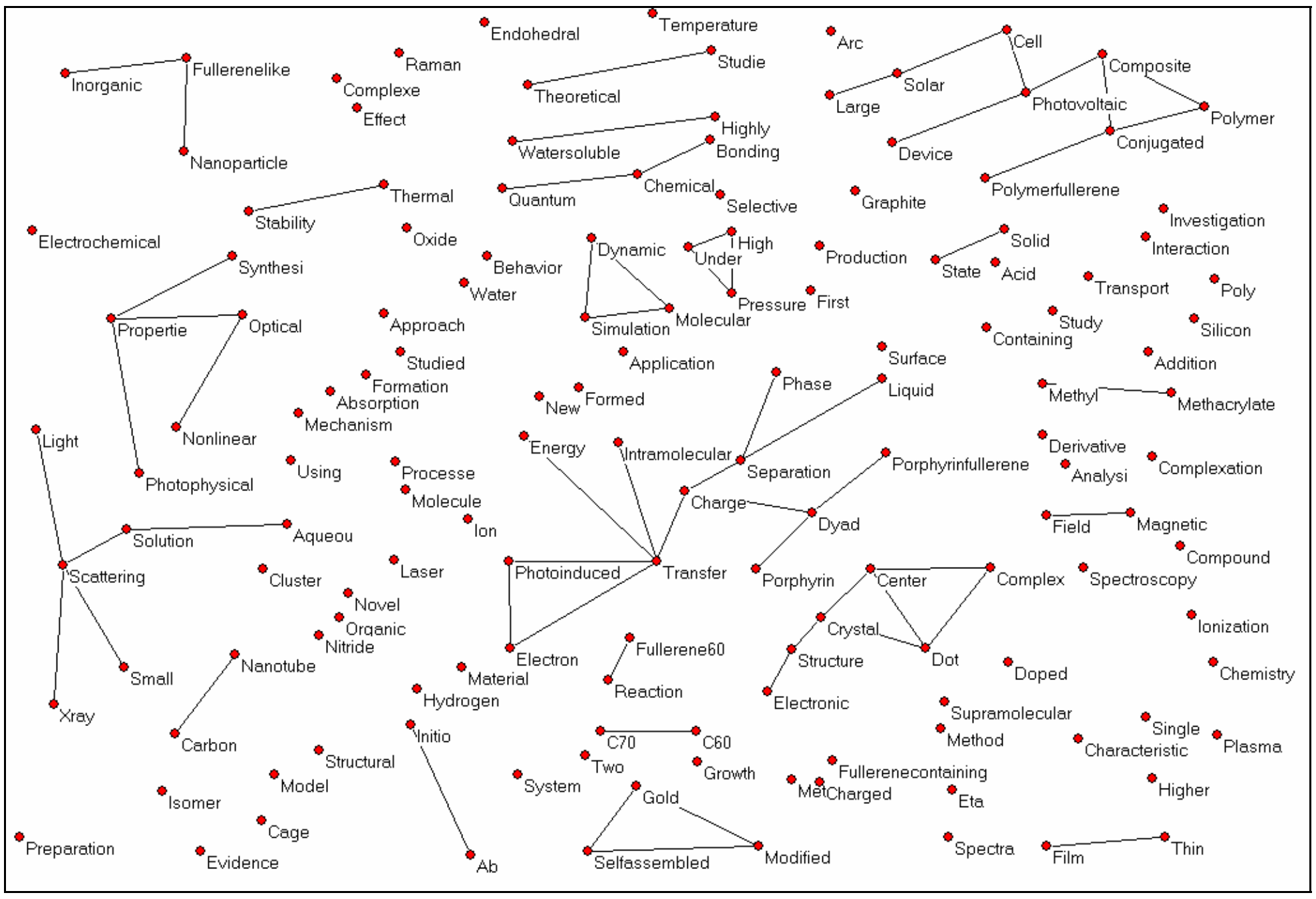

**Figure 9**. Cosine map of the third 2,566 titles in fullerene related documents (cosine ≥ 0.2).

In Figure 9, the semantic map obtained from the analysis of co-occurrences of words in titles exhibits even more sophistication than the previous ones. Some of the clusters from the second set of documents remain, but new nodes and clusters appear. The dyadic relation between carbon and nanotubes demonstrates that the carbon-nanotubes as part of fullerene-related research persist from the earlier documents. The only permanent strong relation that endures from the first set is the link between molecules C-70 and C-60. This link is associated with the production of high purity fullerenes.

*Nanotubes documents*

The exercise was repeated for the 9,672 documents retrieved from the *SCI* for the term nanotube(s) in their titles. Like in the three different sets of fullerene-related documents shown above, the association between carbon and nanotube was present in every set and remained visible using higher thresholds of the cosine (until 0.4) for the last two sets. This indicates the importance of carbon nanotubes in fullerene-related research. As has been mentioned above, carbon nanotubes can be considered as a prolongation of research in fullerenes. The codification of the sets of nanotube-related documents gives insights of further codification of fullerene research when they are no longer the central objective but constitutes underlying knowledge.

For the study of the codification of the concept of "nanotube(s)", three consecutive sets of 3,224 documents were distinguished. The first set (Figure 10) corresponds to documents published from 1992 to 2002. A total of 3,704 words occur in the titles of these documents, and 1,717 occur more than once. For this case 147 words that appeared more than 20 times were analyzed.

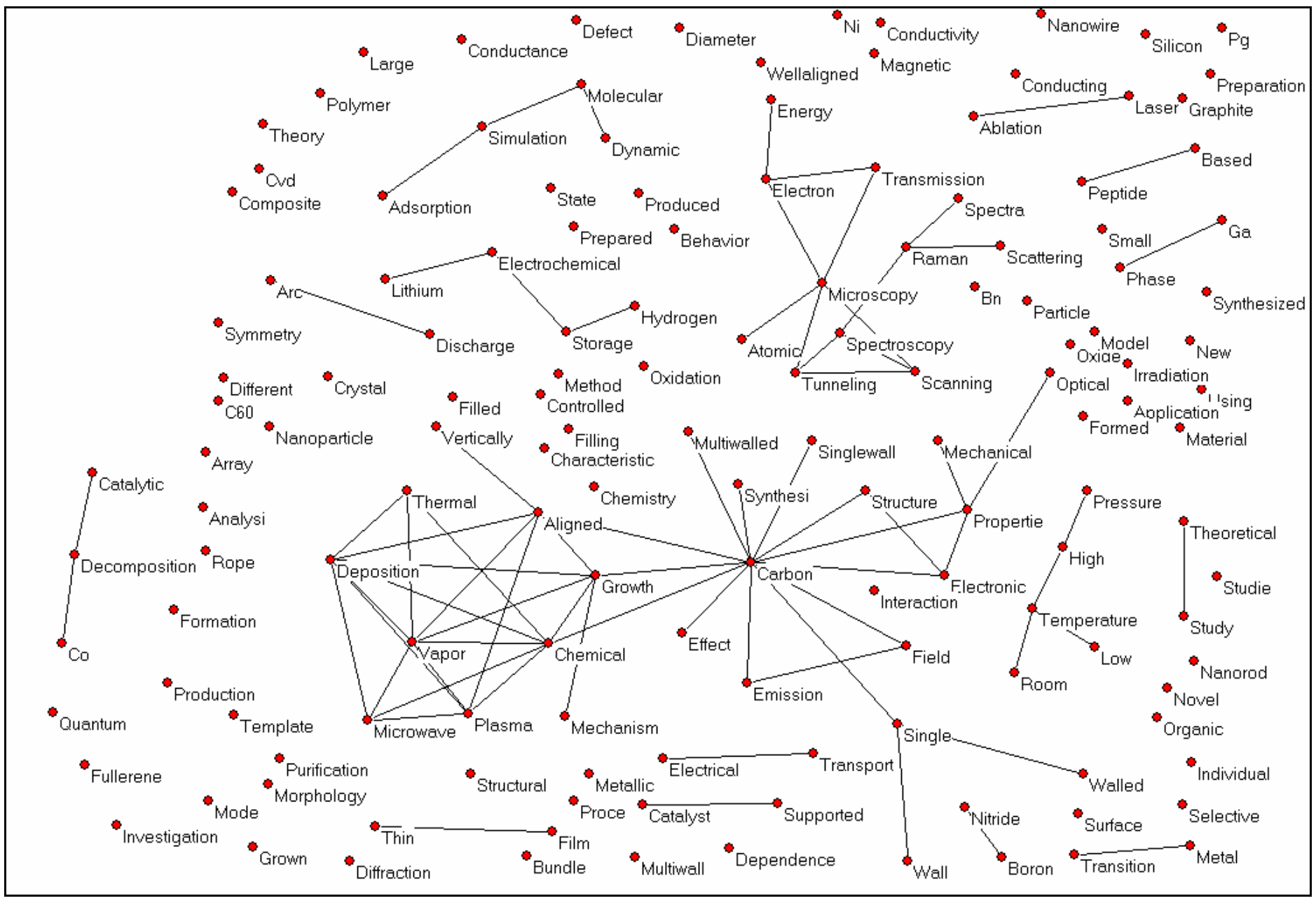

**Figure 10**. Cosine map of the first 3,224 titles in nanotube related documents (cosine ≥ 0.2).

The results of the semantic maps drawn from nanotubes documents agree with findings form prior studies concluding that less codification can be the result of greater variance produced in larger sets of documents (Hellsten & Leydesdorff, 2005). Figure 10 shows this semantic map setting the threshold of the cosine at 0.2 to achieve semantic organization. With this threshold, the network looses its original density and exposes 23 clear clusters in which "carbon" is shown to have a high relevance in the codification of the concept "nanotube".

Figure 11 shows the codification of the second set of titles that corresponds to publications from 2002 to 2004. In them, 4,365 words are present and 1,929 appear at least twice. The analysis was made with 170 words that appeared more than 20 times in the titles of these documents.

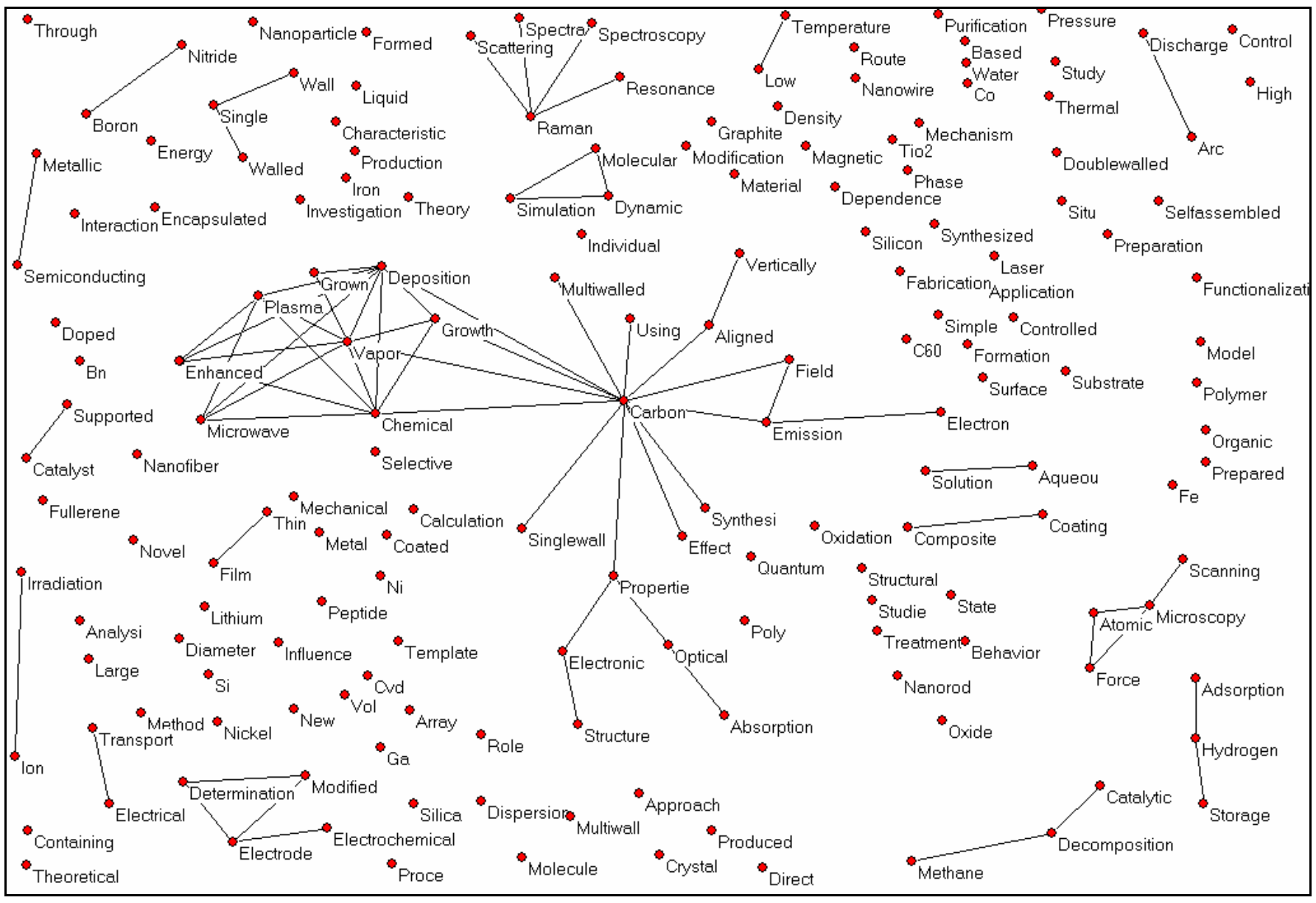

**Figure 11**. Cosine map of the second 3,224 titles in nanotube related documents (cosine ≥ 0.2).

Figure 11 and Figure 10 have similar semantic structures. In Figure 11, clusters that remain from Figure 10 have less nodes participating in them exhibiting less semantic complexity. Carbon still shows a central role in the codification of the carbon nanotubes concept. The reason to leave carbon on the analysis relies in the fact that it was not part of the search term and because not all the documents[5] had carbon in their titles.

The third set (Figure 12) corresponds to documents published between 2004 and 2005. In their titles, a total of 4,811 words appear at least once and 2,099 at least twice. The semantic map is drawn with 169 words that happen more than 20 times.

[5] 73% of the documents have carbon in their titles.

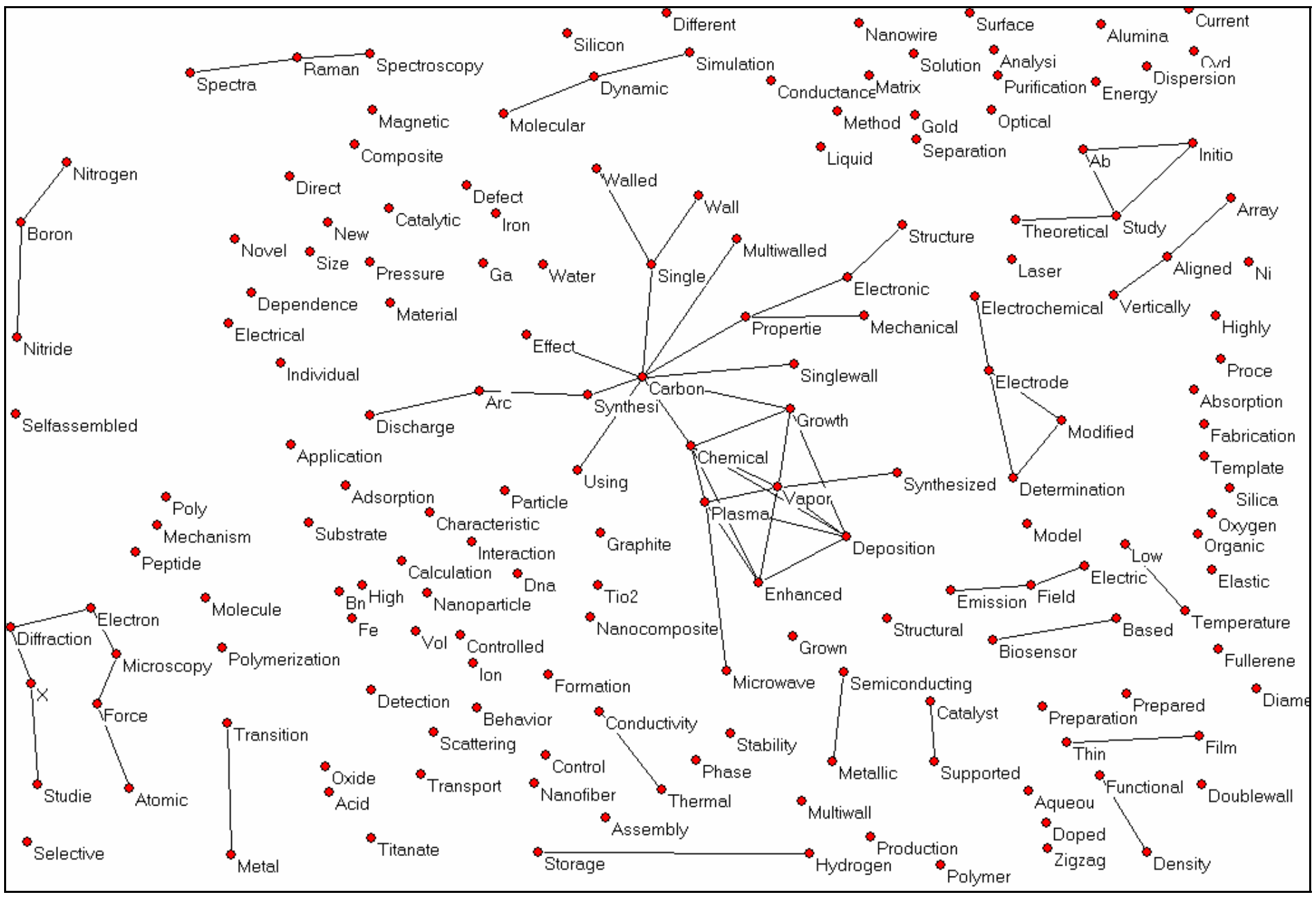

**Figure 12.** Cosine map of the third 3,224 titles in nanotube related documents (cosine ≥ 0.2).

From Figure 12 it can be concluded that less complexity in the clusters has remained in comparison to the first set of documents. Although new clusters appear, the network doesn't display more density as was the case in the evolution of the "fullerene" publications. Carbon remains in this third set of documents the core clustering term even when setting the threshold of the cosine at higher levels.

"Fullerene" is present as an important term in the sets retrieved for "nanotube(s)", although its links are not very strong and disappear when increasing the threshold of the cosine. It holds different relations throughout the three sets. In the first set carbon-fullerenes seems to show the basic relation with fullerene research: nanotubes are fullerene-like carbon structures. In the second, fullerene-quantum-molecules could be more associated to electronic uses; and in the third set, Arc-carbon-fullerene is related to the method to produce the longest most perfectly formed carbon nanotube (Guo *et al.*, 1995)

*4.2.2 Codification at different domains (patents)*

Until now, meaning codification has been analyzed in the scientific domain giving attention to possible changes through time. But the codification of meaning is expected to vary when communication takes place at different domains (Leydesdorff & Hellsten, 2005). The processes in each of the domains and their different purposes in the use and treatment of knowledge result in different functions and meaning of words.

Information was retrieved from the USPTO (www.uspto.org) for all patents applications in the United States that had fullerene(s) or nanotube(s) in their titles.[6] The potential applications of both fullerenes and nanotubes in the development of new materials were expected to be evident in the codification of the concepts in the titles of these documents.

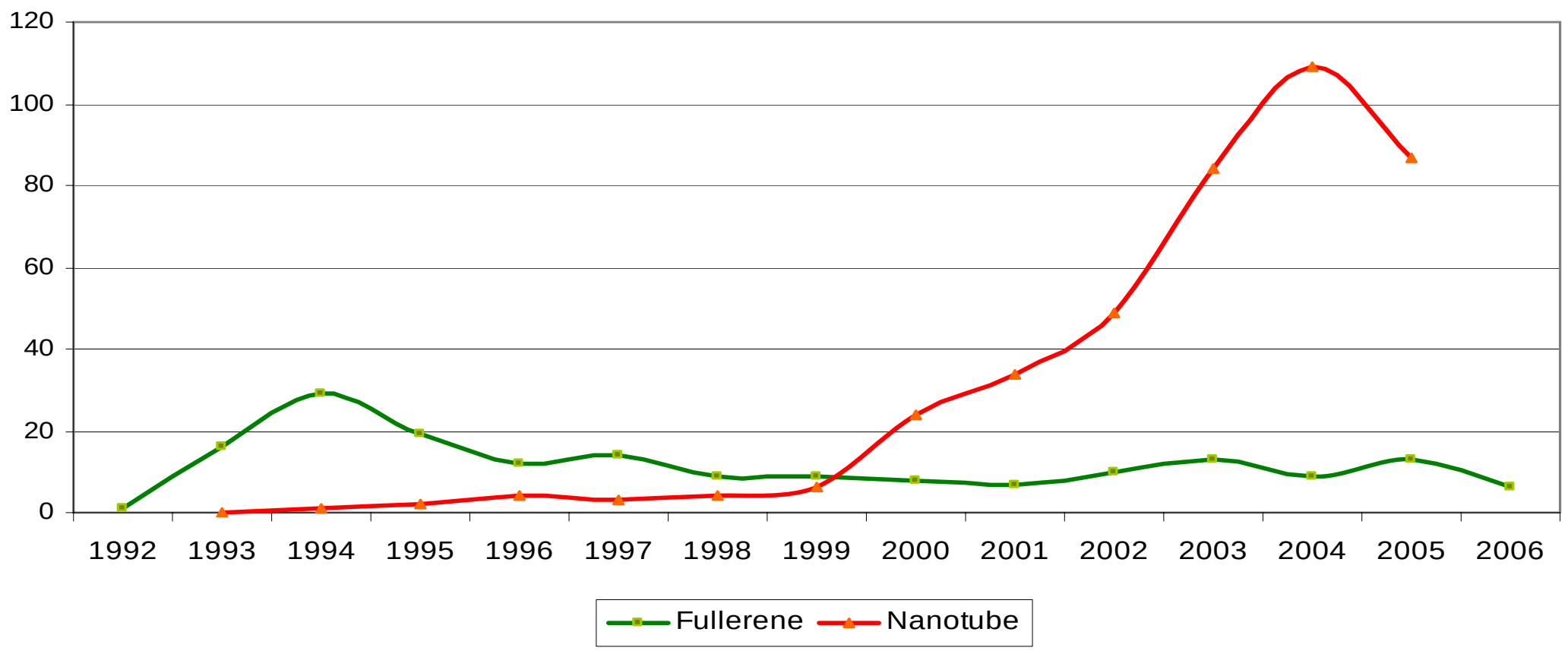

**Figure 13.** Number of patent applications in USPTO with “fullerene(s)” or “nanotube(s)” in their titles over time.

Figure 13 resembles Figure 1 in the tendency of the yearly increase of documents. Patents applications for fullerenes started earlier and had a higher growing rate in earlier stages. Properties of carbon nanotubes evidenced more potential applications than just fullerenes; applications of nanotubes adopted an exponential growing rate (as the documents published) while for fullerenes they stabilized with time. An early growth of fullerenes-applications, its stabilization and a further ‘boom’ in nanotubes-applications can respond to an innovation process characterized by an early science-push followed by market-push (Schmoch, 1997). What is evident from both Figure 1 and Figure 13, is that even though nanotubes is a spin-off effect of research in fullerenes, it differentiates and

[6] ttl/fullerene$ ; ttl/nanotube$

forms its own research front. For the case of patents applications, just 4 documents have both the word nanotubes and fullerenes in their title.

*Fullerene patent documents*

For the case of fullerene, the exercise retrieved a total of 175 patents, with “fullerene(s)” in their titles, in which 411 words happened. For this analysis 117 words that occurred more than twice were used. The semantic map of the title words of these 175 patent documents is illustrated in Figure 14.

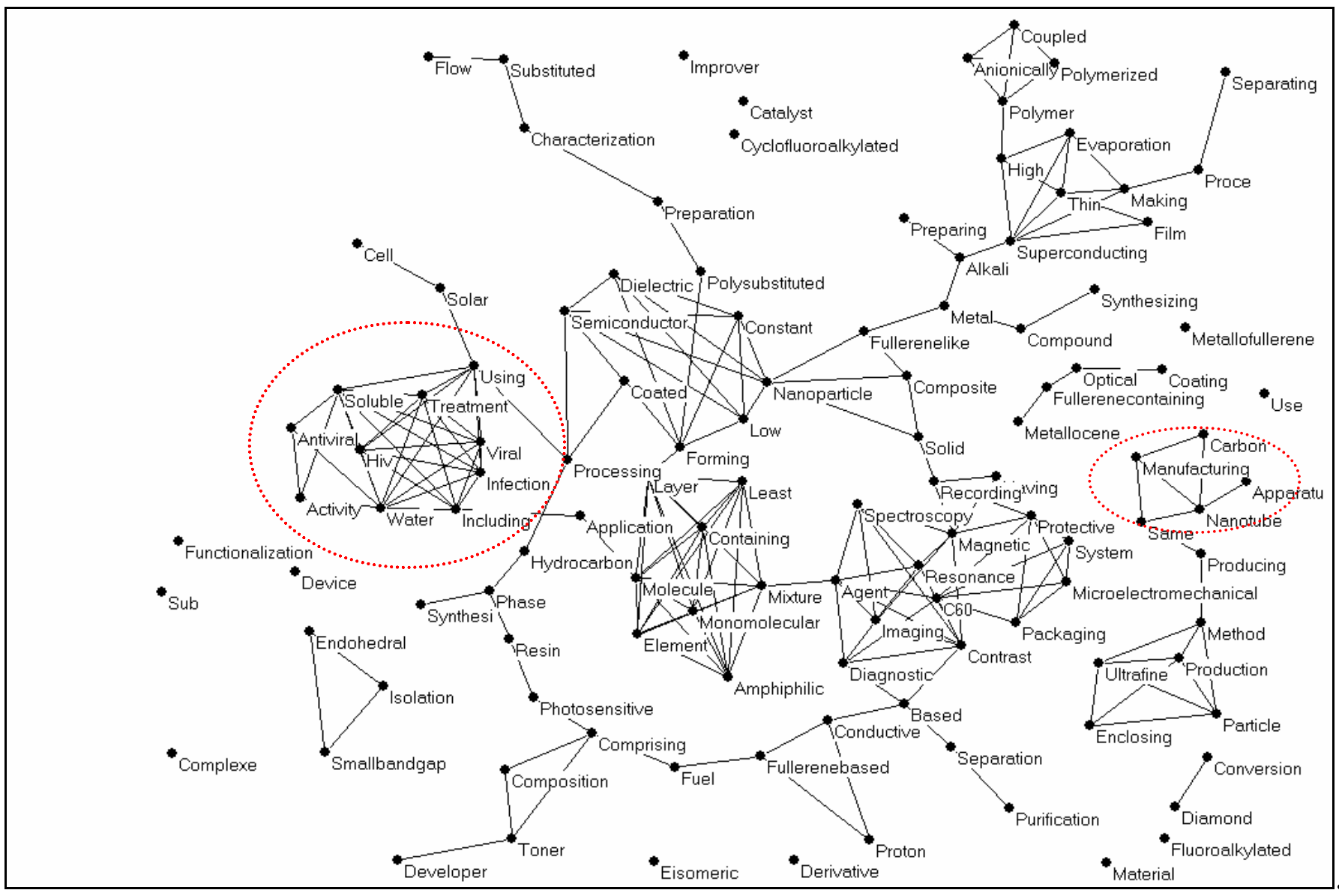

**Figure 14.** Cosine map of the titles in fullerene related patent documents (cosine ≥ 0.3).

From Figure 14, the codification of the meaning of fullerene does not seem to be strongly codified in relation to its envisaged applications and differently from codifications at the scientific domain. In fact just the cluster that is formed with HIV-Viral-Infection-Treatment-Soluble appears as a complete novelty from the frequent words in the titles of scientific communication.

The rest of the clusters seem to be related to processes of preparation and qualities and characteristics of fullerenes; words all frequent in the scientific communication. The cluster formed by the words carbon-manufacturing-nanotube-same-apparatus, should be related to the production of carbon nanotubes. A further analysis should be based on the relation between the patents and its references to examine the scientific-base in fullerenes and nanotubes patent applications.

*Nanotubes patent documents*

For the case of nanotubes, 407 patents with “nanotube(s)” in their titles were retrieved. As illustrated in Figure 13, nanotube applications manifest later in the patents documents, but they soon experiment increasing growing rates. The apparent decrease of this tendency for 2006 is consequence of the moment of the query.

In these titles, a total of 775 words occurred and 301 had a higher frequency than two. The analysis was made with 217 of these words that appeared more than twice, removing propositions, the term nanotubes and taking only the singular form of the words.

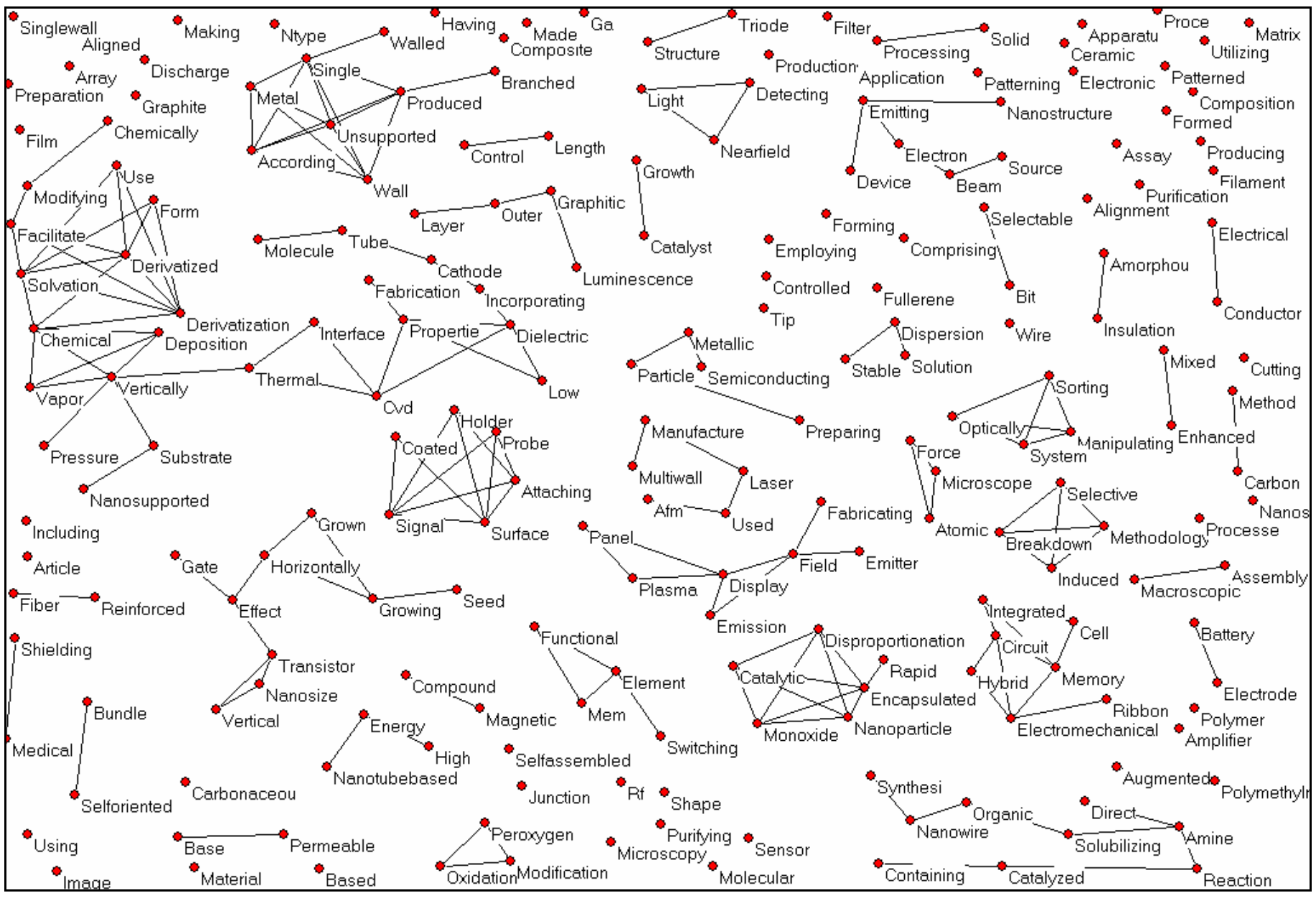

**Figure 15.** Cosine map of the titles in nanotube related patent documents (cosine ≥ 0.4).

The results from the analysis of the titles in the nanotube patents in Figure 15 show more relation to its expected applications in comparison to fullerene patent documents. Clustering words such as emitter, plasma, signal, fiber, medical shielding, nanosize transistor, growing seed, high energy, emitting device, electrical conductor, cell, memory, circuits, battery electrode, nanowire, seem to be more related to potential applications.

Nevertheless, the permanence, when increasing the threshold of the cosine, of some clusters formed with words such as cvd[7], methodology, discharge, chemical, vapor, deposition are very related to nanotube production. This indicates that the tension to patent nanotube producing methods continues to be very important in relation to its further applications.

"Building-blocks" properties of nanotubes expose a broad range of potential commercial applications as response to further research in the field. "Nanotech researchers worldwide are steadily filing patents in the hopes of creating 'tollbooths' for future products incorporating nanomaterials" (Lux Research Report, 2005). The semantic maps obtained for patent documents with "fullerene(s)" and "nanotube(s)" in their titles; seem to reveal this excessive patenting situation. Indications of consumer applications from the semantic maps are very limited and word-clusters are more related to the production and properties of the molecules. If this behavior continues, the possibility to apply fullerenes and nanotubes in consumer oriented devices might be limited as the costs of licensing can become inefficiently high. (*Ibid.*)

### *4.3 Citation behavior*

For the last part of the analysis the software *HistCite* was used to establish the citing relations inside the fullerene and nanotube related documents retrieved from the *SCI.* This software feeds from files obtained from queries to the Web of Science and highlights different aspects of a set of documents such as citations inside the set, most relevant authors or institutions, links inside the set, missing links and most cited nodes among others.

The importance of this particular software for this part of the analysis is that it creates a citation index for the documents retrieved. Chronological citation maps are automatically generated limiting the citations to the desired set of documents. This facilitates historical reconstruction of scientific developments. This way, differences in the citing dynamics inside each of the sets (fullerene and nanotubes) can be associated with differences in the emergence and stabilization of research fronts.

[7] Chemical Vapor Deposition

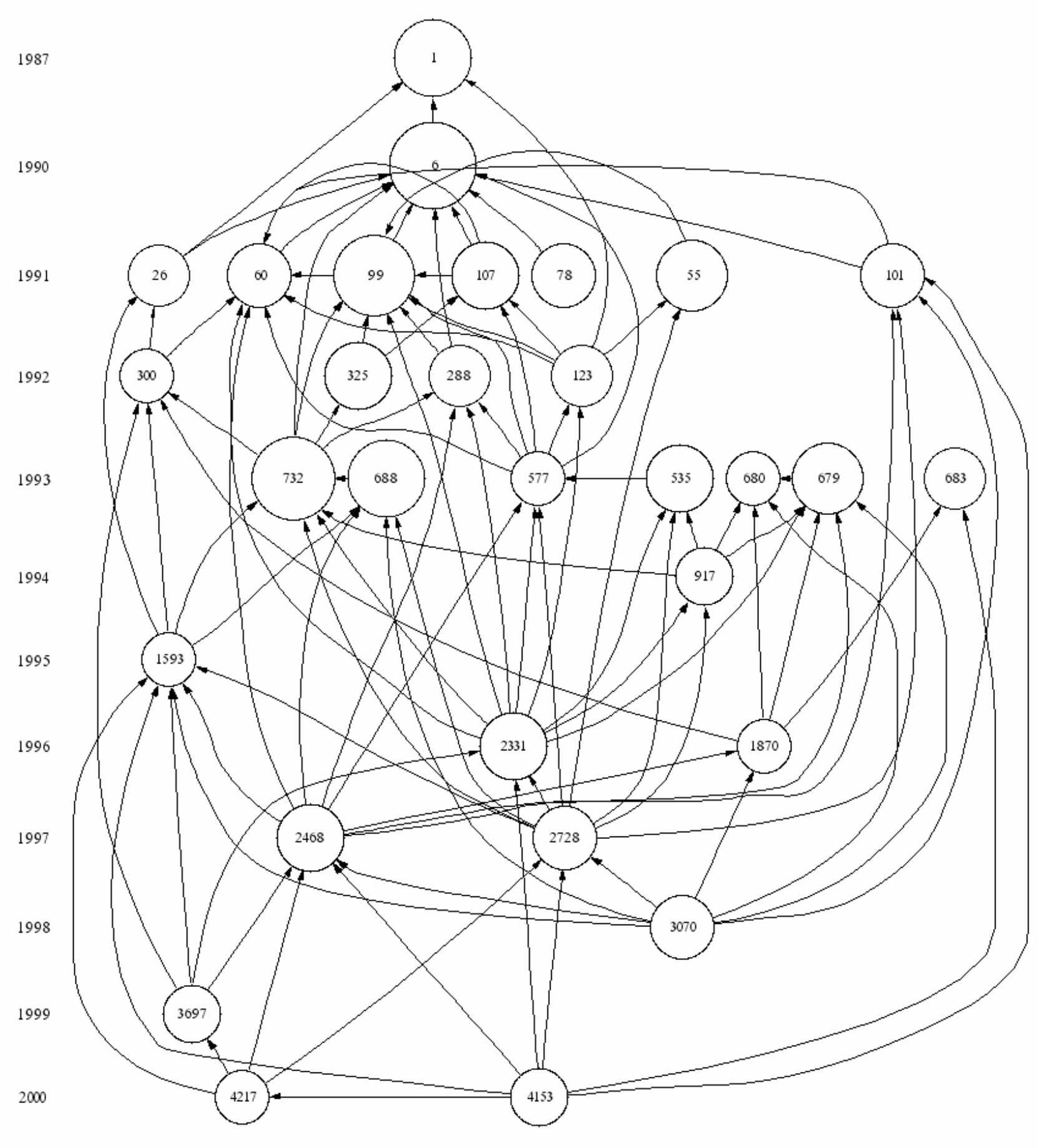

1 KROTO HW, 1987, NATURE, V329, P529
6 TAYLOR R, 1990, J CHEM SOC CHEM COMMUN, P1423
26 HARE JP, 1991, CHEM PHYS LETT, V177, P394
55 CHAI Y, 1991, J PHYS CHEM, V95, P7564
60 ALLEMAND PM, 1991, J AMER CHEM SOC, V113, P1050
78 HOWARD JB, 1991, NATURE, V352, P139
99 DIEDERICH F, 1991, SCIENCE, V252, P548
101 ALLEMAND PM, 1991, SCIENCE, V253, P301
107 DIEDERICH F, 1991, SCIENCE, V254, P1768
123 DIEDERICH F, 1992, ACCOUNT CHEM RES, V25, P119
288 CREEGAN KM, 1992, J AMER CHEM SOC, V114, P1103
300 ANDERSSON T, 1992, J CHEM SOC CHEM COMMUN, P604
325 KIKUCHI K, 1992, NATURE, V357, P142
535 BINGEL C, 1993, CHEM BER-RECL, V126, P1957
577 ISAACS L, 1993, HELV CHIM ACTA, V76, P1231
679 FRIEDMAN SH, 1993, J AMER CHEM SOC, V115, P6506
680 SIJBESMA R, 1993, J AMER CHEM SOC, V115, P6510
683 TOKUYAMA H, 1993, J AMER CHEM SOC, V115, P7918

| | |
|---|---|
| 688 | MAGGINI M, 1993, J AMER CHEM SOC, V115, P9798 |
| 732 | TAYLOR R, 1993, NATURE, V363, P685 |
| 917 | HIRSCH A, 1994, ANGEW CHEM INT ED, V33, P437 |
| 1593 | WILLIAMS RM, 1995, J AMER CHEM SOC, V117, P4093 |
| 1870 | Jensen AW, 1996, BIOORGAN MED CHEM, V4, P767 |
| 2331 | Diederich F, 1996, SCIENCE, V271, P317 |
| 2468 | Imahori H, 1997, ADVAN MATER, V9, P537 |
| 2728 | Prato M, 1997, J MATER CHEM, V7, P1097 |
| 3070 | Prato M, 1998, ACCOUNT CHEM RES, V31, P519 |
| 3697 | Diederich F, 1999, CHEM SOC REV, V28, P263 |
| 4153 | Guldi DM, 2000, ACCOUNT CHEM RES, V33, P695 |
| 4217 | Guldi DM, 2000, CHEM COMMUN, P321 |

**Figure 16.** Citing relations among documents with "fullerene(s)" in their titles, *HistCite*

For the 7,962 documents that have fullerene in their titles, Figure 16 shows the relations between the 30 most cited documents inside the set[8]. The size of the nodes indicates the amount of citations it gets in the set. There are 3,050 documents that are not cited by other documents inside the set, while 1,237 documents do not cite any of the documents in the set.

One of the most cited articles inside the set was "C60: Buckministerfullerene" (Kroto, 1985). This document was not included in the set. A further analysis including the 408 documents from the *SCI* with Buckministerfullerene* in their titles might be appropriate.

[8] Inside the set, 63,026 references were outside the set of documents. Two references are worth mentioning: "C60 Buckministerfullerene" published in *Nature* in 1985 and referenced 1,495 times in the set and "Solid C60: A new form of carbon" published in Nature in 1990 and being referenced 1,410 times

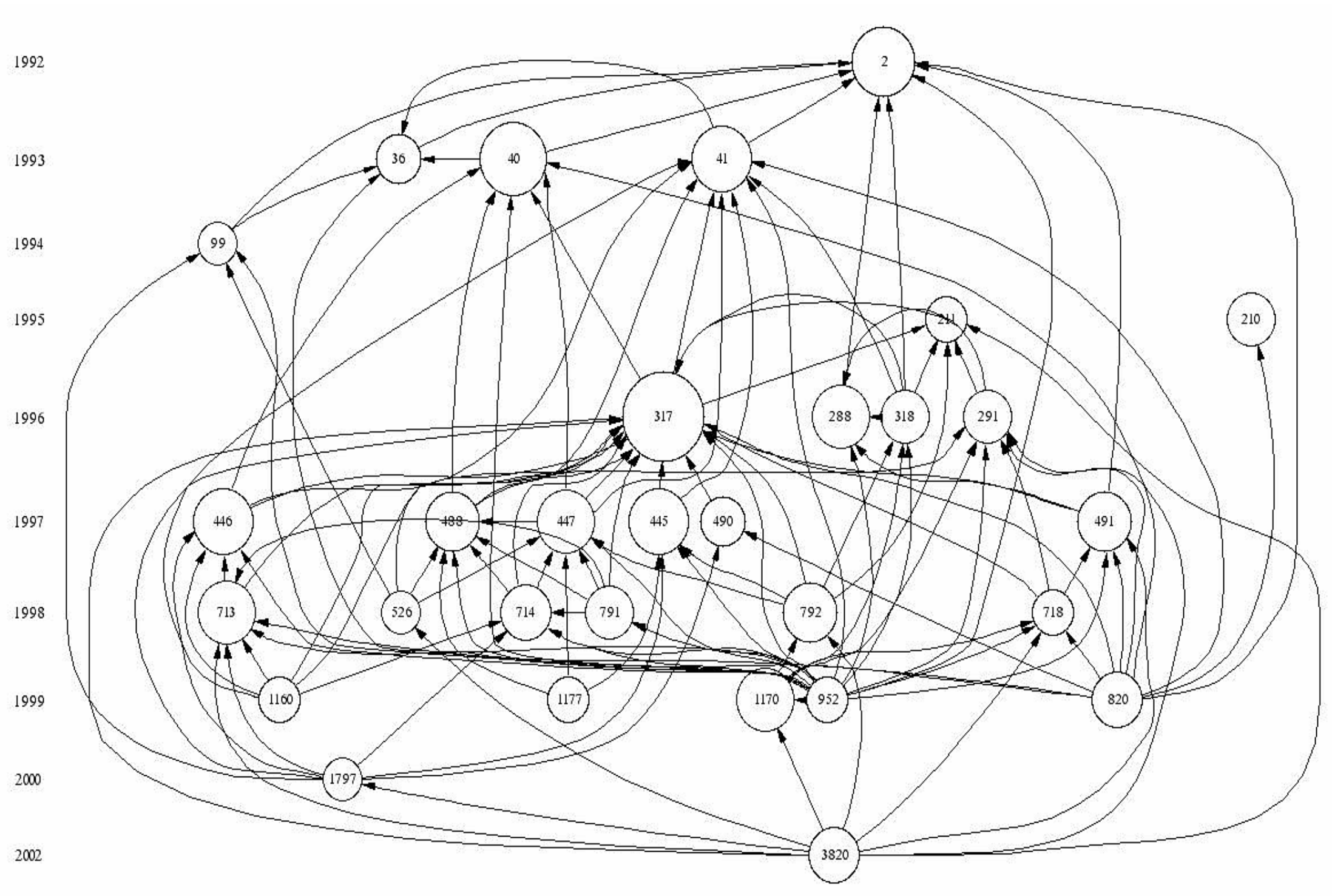

2 EBBESEN TW, 1992, NATURE, V358, P220
36 AJAYAN PM, 1993, NATURE, V361, P333
40 IIJIMA S, 1993, NATURE, V363, P603
41 BETHUNE DS, 1993, NATURE, V363, P605
99 TSANG SC, 1994, NATURE, V372, P159
112 BLASE X, 1994, PHYS REV LETT, V72, P1878
163 GUO T, 1995, CHEM PHYS LETT, V243, P49
210 CHOPRA NG, 1995, SCIENCE, V269, P966
211 RINZLER AG, 1995, SCIENCE, V269, P1550
288 Treacy MMJ, 1996, NATURE, V381, P678
291 Dai HJ, 1996, NATURE, V384, P147
317 Thess A, 1996, SCIENCE, V273, P483
318 Li WZ, 1996, SCIENCE, V274, P1701
445 Dillon AC, 1997, NATURE, V386, P377
446 Tans SJ, 1997, NATURE, V386, P474
447 Journet C, 1997, NATURE, V388, P756
488 Rao AM, 1997, SCIENCE, V275, P187
490 Bockrath M, 1997, SCIENCE, V275, P1922
491 Wong EW, 1997, SCIENCE, V277, P1971
526 Rinzler AG, 1998, APPL PHYS A-MAT SCI PROCESS, V67, P29
713 Wildoer JWG, 1998, NATURE, V391, P59
714 Odom TW, 1998, NATURE, V391, P62
765 Bandow S, 1998, PHYS REV LETT, V80, P3779
791 Chen J, 1998, SCIENCE, V282, P95
792 Ren ZF, 1998, SCIENCE, V282, P1105
950 Nikolaev P, 1999, CHEM PHYS LETT, V313, P91

952 Ajayan PM, 1999, CHEM REV, V99, P1787
1170 Fan SS, 1999, SCIENCE, V283, P512
1214 Kataura H, 1999, SYNTHET METAL, V103, P2555
3820 Baughman RH, 2002, SCIENCE, V297, P787

**Figure 17.** Citing relations among nanotube-related documents, *HistCite*

In Figure 17, the citing relations inside the set of nanotubes documents are depicted. The total number of documents that do not cite any other document in the set is 1,035 while, 4,517 documents are not cited by any other document of the set.

The citation dynamics inside each of the sets illustrates different behaviors. For the set of fullerene, a clear vertical hierarchy illustrates that early documents continue being very relevant for the development of the field. For the case of nanotubes, the relations are more horizontally shaped and there is not one single document with an extraordinary citation score. The importance of citations of recent documents in the set of nanotubes is typical of the development of research fronts (Chen, 2006).

If nanotubes research have characteristic of a research fronts, fullerene research could be forming the intellectual base of this new development. Measuring the integration of knowledge acquired in fullerene research in the further development of a more specialized research front in nanotubes is possible analyzing both sets of documents together.

In Figure 18, citing relations of both sets are illustrated. Inside the set, 2,154 do not cite any of the documents of the set and 7,377 are not cited by other documents in the set.

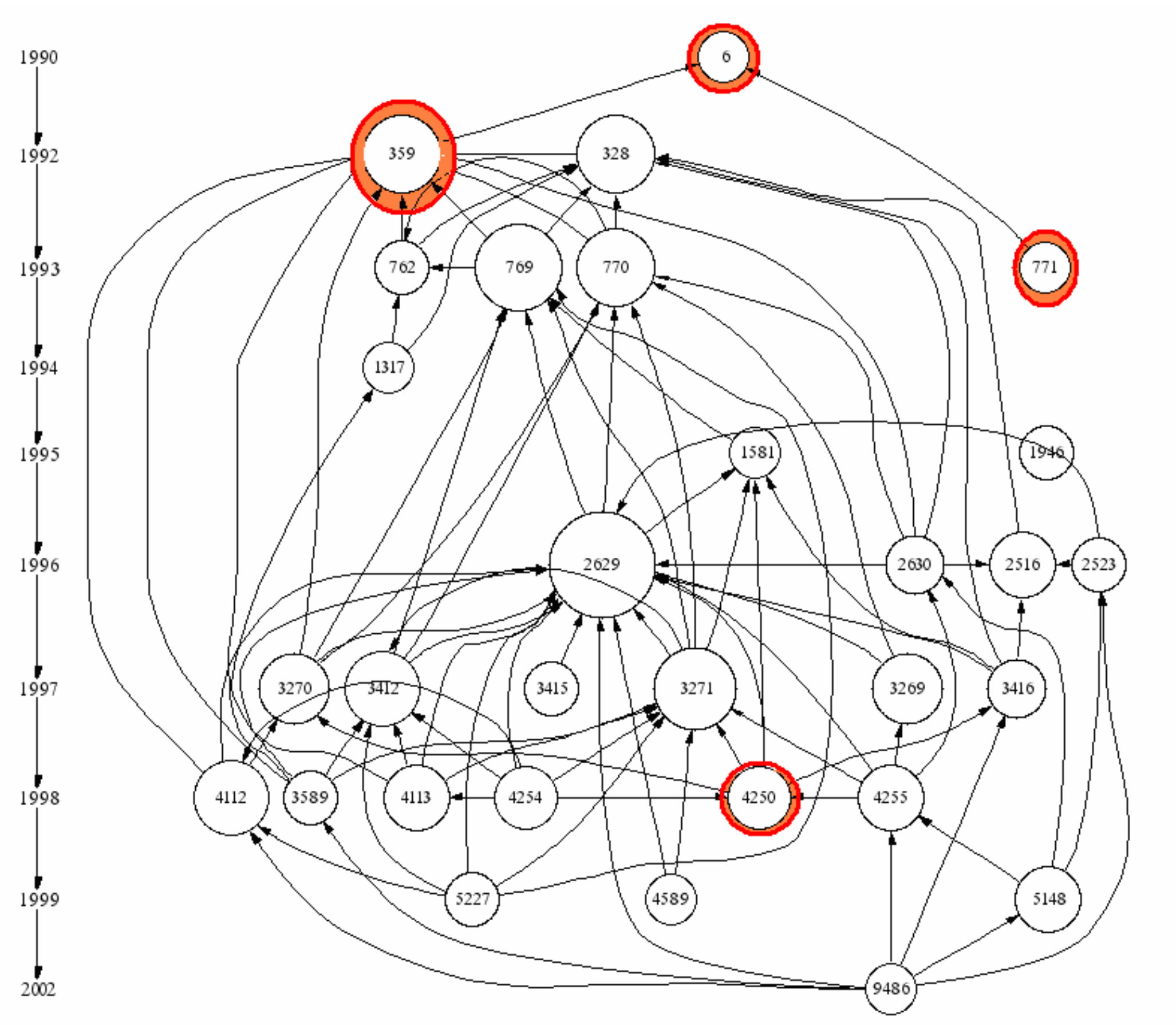

6 TAYLOR R, 1990, J CHEM SOC CHEM COMMUN, P1423
328 EBBESEN TW, 1992, NATURE, V358, P220
359 MINTMIRE JW, 1992, PHYS REV LETT, V68, P631
762 AJAYAN PM, 1993, NATURE, V361, P333
769 IIJIMA S, 1993, NATURE, V363, P603
770 BETHUNE DS, 1993, NATURE, V363, P605
771 TAYLOR R, 1993, NATURE, V363, P685
1317 TSANG SC, 1994, NATURE, V372, P159
1581 GUO T, 1995, CHEM PHYS LETT, V243, P49
1946 CHOPRA NG, 1995, SCIENCE, V269, P966
2516 Treacy MMJ, 1996, NATURE, V381, P678
2523 Dai HJ, 1996, NATURE, V384, P147
2629 Thess A, 1996, SCIENCE, V273, P483
2630 Li WZ, 1996, SCIENCE, V274, P1701
3269 Dillon AC, 1997, NATURE, V386, P377
3270 Tans SJ, 1997, NATURE, V386, P474
3271 Journet C, 1997, NATURE, V388, P756
3412 Rao AM, 1997, SCIENCE, V275, P187
3415 Bockrath M, 1997, SCIENCE, V275, P1922
3416 Wong EW, 1997, SCIENCE, V277, P1971
3589 Rinzler AG, 1998, APPL PHYS A-MAT SCI PROCESS, V67, P29

4112 Wildoer JWG, 1998, NATURE, V391, P59
4113 Odom TW, 1998, NATURE, V391, P62
4250 Liu J, 1998, SCIENCE, V280, P1253
4254 Chen J, 1998, SCIENCE, V282, P95
4255 Ren ZF, 1998, SCIENCE, V282, P1105
4589 Nikolaev P, 1999, CHEM PHYS LETT, V313, P91
5148 Fan SS, 1999, SCIENCE, V283, P512
5227 Kataura H, 1999, SYNTHET METAL, V103, P2555
9486 Baughman RH, 2002, SCIENCE, V297, P787

**Figure 19.** Citing relations among nanotube- and fullerene-related documents, *HistCite*. Highlighted nodes denote documents with the word "fullerene(s)" in their titles.

Although nanotubes have been recognized as a type of one-dimensional fullerene, and their discovery can be understood as a trigger effect of research in fullerene, the amount of documents published with nanotube in their title and the participation of countries in their production can be embedded in a shift of effort and attention in the different research fronts.

## Conclusions

Network analysis proves useful to understand the dynamic process of knowledge production as a further codification of information, in this text as a consequence of new discoveries. In the journal space, the emergence of a new journal as a consequence of discoveries marks an important breakthrough. For this particular case, *Fullerene, Science and Technology* publishes an important share of documents related to fullerenes but looses importance in the distribution of communication in nanotubes (only 93 of 840 documents titles in the journal contain the word "nanotube(s)").

For the semantic codification of meaning in time, while "carbon-nanotubes" constitute a dyadic relation for all the time-differentiated sets of fullerene documents, "fullerene" has weak patterns of association in the sets of nanotube documents. Relations among words were different in each of the sets. While for fullerenes the network became denser with more complex clusters, in nanotubes the semantic codifications in time did not acquire more complexity. This seems to prove that there are underlying processes that generate structures in the communication: some provide variation and others are more functionally oriented to create structure in the communication.

Although there were some differences in the codifications in different domains, the most frequent words were similar in scientific and patent documents. The hype in patenting fullerene and nanotube related processes can affect the semantic codification of its commercial applications.

The citation relations showed that “fullerene” documents are among the most cited documents in the development of nanotube research. A deeper analysis of references and most cited documents of the set must be done to make further remarks about the constitution of fullerene research as the intellectual base of the nanotube research front. For the case of fullerenes, the most frequently cited articles are not the more recent ones as could be expected of a nanoscience field of research.

The different network analysis seems to converge in the conclusion that even though fullerene and nanotube research are related, it seems that nanotube research has acquired its own dynamics soon after its discovery and has been disconnected from fullerene research. This is confirmed by the fact that there is a different distribution of authoring countries in each of the sets of documents. United States and Japan have an important input in both fronts, but actors such as Switzerland, Taiwan and South Korea have stronger participation in research in nanotubes than in fullerenes. Countries like China, Germany, England, France and Italy have a relevant participation in the production of fullerene documents, but not significant for the case of nanotubes.

One of the possible reason why fullerene and nanotube research drift and constitute each its own dynamics and institutions is that although they are both nanometer carbon structures, the exceptionality of the molecular nature of the nanotubes in terms of their high length-to-width aspect ratio has captured the attention of further research as specialized and large-scale applications are growing constantly (Colbert, 2003).

**Further work**

Knowledge has become increasingly central to the progress of society and our daily lives. Industrial economies are being targeted to become more knowledge-based. This justifies the importance of finding ways to operationalize the knowledge-base, to observe and measure the fluxes of information that form its foundations, and to understand issues concerning the processes behind the production and application of knowledge.

Our purpose in this paper has been to understand the emergence of knowledge as a consequence of new discoveries and its disrupting effects in the structures of scientific communications. To do this, we have concentrated on the relations among the codes of communication, and the changes of meaning codification in relation to time and domain. Our first assumption has been that knowledge is socially constructed through communication.

We consider time as crucial in the process of knowledge construction and application. New discoveries constitute new information that acts generate variation in the communication, but this can only be observed as time-dependent variation. New information can become knowledge with the introduction of time as a further degree of freedom through which the disrupted communication re-organizes. Different properties and measures of the networks could give some hindsight into the state of development of scientific disciplines, topics of research and research fronts, as well as the integration of knowledge to existing knowledge bases.

In this study we have worked with the codification of two concepts related to nanoscience, an emerging field of the sciences. The analysis has been to understand the changes in scientific communication. Further steps will consider knowledge-based innovation to measure their scientific base and analyze codification processes at the interfaces as well as concepts related to medicine and the social sciences.

**Acknowledgment**

We wish to thank Eugene Garfield and Alexander Pudovkin for allowing us to use HistCite™ for the evaluation.